\documentclass{iopjournal}

\usepackage{amsmath}
\usepackage{hyperref}
\usepackage{mathrsfs}
\usepackage{bm}
\usepackage{cancel}
\usepackage{dsfont}
\usepackage{caption}
\usepackage{subcaption} 
\usepackage{times}
\usepackage{lipsum}
\usepackage[normalem]{ulem}
\usepackage{amsmath,amsfonts,amssymb,latexsym,amsthm,bbm}
\usepackage{graphicx,epsf}
\usepackage[toc,page,header]{appendix}
\usepackage{minitoc}
\usepackage[numbers,sort&compress]{natbib}
\usepackage{color}
\newcommand{\be}{\begin{equation}}
\newcommand{\ee}{\end{equation}}
\newcommand{\ba}{\begin{eqnarray}}
\newcommand{\ea}{\end{eqnarray}}
\newcommand{\ignore}[1]{}

\usepackage{tikz}
\usetikzlibrary{quantikz2,shapes,arrows.meta}

\usepackage{algorithm}
\usepackage{algpseudocode}

\newcommand{\PSTAB}{\mathsf{PSTAB}}

\newcommand{\norm}[1]{\left\lVert #1 \right\rVert}
\newcommand{\abs}[1]{\left| #1 \right|}

\newcommand{\hi}{\mathcal{H}}
\newcommand{\id}{\mathbbm{1}}

\DeclareMathOperator{\diag}{diag}

\newcommand{\1}{\mathbbm{1}}

\newtheorem{definition}{Definition}

\begin{document}

\title{Van Hove singularities in stabilizer entropy densities}

\author{Daniele Iannotti$^{1,2,*}$\orcid{0009-0009-0738-5998}, Lorenzo Campos Venuti$^{2,3}$\orcid{0000-0002-0217-6101} and Alioscia Hamma$^{1,2,3}$\orcid{0000-0003-0662-719X}}

\affil{$^1$Scuola Superiore Meridionale, Largo S. Marcellino 10, 80138 Napoli, Italy}
\affil{$^2$Istituto Nazionale di Fisica Nucleare (INFN) Sezione di Napoli, Napoli, Italy}
\affil{$^3$Università degli Studi di Napoli Federico II , Dipartimento di Fisica Ettore Pancini, Napoli, Italy}

\email{* d.iannotti@ssmeridionale.it}


\begin{abstract}
The probability distribution of a measure of non-stabilizerness, also known as magic, is investigated for Haar-random pure quantum states. Focusing on the stabilizer Rényi entropies, the associated probability density functions (PDFs) are found to display distinct non-analytic features analogous to Van Hove singularities in condensed matter systems. For a single qubit, the stabilizer purity exhibits a logarithmic divergence at a critical value corresponding to a saddle point on the Bloch sphere. This divergence occurs at the $|H\rangle$-magic states, which hence can be identified as states for which the density of non-stabilizerness in the Hilbert space is infinite. 
An exact expression for the PDF is derived for the case $\alpha = 2$, with analytical predictions confirmed by numerical simulations. The logarithmic divergence disappears for dimensions $d \ge 3$, in agreement with the behavior of ordinary Van Hove singularities on flat manifolds.
In addition, it is shown that, for qubits systems, the linear stabilizer entropy is directly related to the \emph{partial incompatibility} of quantum measurements, one of the defining properties of quantum mechanics, at the basis of Stern-Gerlach experiments. 
\end{abstract}

\section{Introduction}
Characterizing features of quantum resources~\cite{gour2024resources,chitambar2019QuantumResourceTheories,Gour_2025} has been an important and challenging issue since the development of quantum computation in the past years, beyond, of course, the fundamental reasons connected to understand quantum mechanics. 
In this context, entanglement played an enormous role as its power for quantum algorithms was made clearer, as well as being able to distinguish quantum mechanics from classical theories~\cite{horodecki2009QuantumEntanglement,huang2025vastworldquantumadvantage}.
Over the past twenty years, another important resource, non-stabilizerness, or \emph{magic}, has taken shape, rooted in the pioneering work of Bravyi and Kitaev~\cite{bravyi2005UniversalQuantumComputation}. They demonstrated that non-Clifford gates, essential for universal quantum computation, can be realized by augmenting stabilizer operations (i.e., Clifford gates and projective measurements) with special nonstabilizer, or “magic” states. 
The fundamental insight is that while stabilizer operations can typically be implemented in a fault-tolerant manner, often via transversal methods in quantum error-correcting codes~\cite{campbell2010BoundStatesMagic}, non-Clifford gates remain considerably more challenging to protect against noise~\cite{PhysRevLett.102.110502,turkeshi2024coherent}. 
Additionally, magic plays an important role in understanding the classical simulability of quantum circuits. Namely, circuits consisting only of stabilizer operations can be efficiently simulated on a classical computer~\cite{gottesman1998heisenberg}, whereas allowing for the presence of nonstabilizer states drastically increases  the simulation complexity~\cite{aaronson2004ImprovedSimulationStabilizer}.

The interest in understanding the main features of magic~\cite{veitch2014ResourceTheoryStabilizer,Heimendahl_2022,zhu2016clifford,Iannotti_2025,3ttm-vhdt,chitambar2019QuantumResourceTheories,aditya2025mpembaeffectsquantumcomplexity,Haug_2023,leone2025noncliffordcostrandomunitaries,bittel2025completetheorycliffordcommutant,hou2025stabilizer} has grown rapidly, since its first insights. Spanning from quantum circuit~\cite{scocco_rise_2025,magni_anticoncentration_2025,magni_quantum_2025,mittal_quantum_2025,varikuti_impact_2025,Paviglianiti:2025uhy,lóio2025quantumstatedesignsmagic,p8dn-glcw,varikuti2025impactcliffordoperationsnonstabilizing,zhang2024quantummagicdynamicsrandom}, condensed matter and many-body physics~\cite{oliviero_magic-state_2022,haug_quantifying_2023,passarelli_nonstabilizerness_2024,passarelli_chaos_2025,rattacaso_stabilizer_2023,tirrito_anticoncentration_2024,tirrito_universal_2025,tirrito_magic_2025,russomanno_efficient_2025,passarelli_nonstabilizerness_2025,sticlet_non-stabilizerness_2025,viscardi_interplay_2025,jasser_stabilizer_2025,collura_quantum_2025,falcao_nonstabilizerness_2025,brokemeier_quantum_2025,tarabunga2025efficientmutualmagicmagic,PhysRevB.110.045101,Sarkar_2020,Liu_2022,PhysRevB.111.054301,zhang2025stabilizerrenyientropytransition,moca2025nonstabilizernessgenerationmultiparticlequantum,bera2025nonstabilizernesssachdevyekitaevmodel,Korbany_2025,PhysRevB.107.035148,PhysRevLett.133.010601,sticlet2025nonstabilizernessopenxxzspin,mx8t-l4hf,moca2025nonstabilizernessdiagnosticcriticalityexceptional}, to nuclear physics~\cite{sarkis_are_2025,savage_quantum_2025,robin_magic_2024}, particle physics and lattice gauge theories~\cite{busoni_emergent_2025,illa_dynamical_2025,white_magic_2024,aoude_probing_2025,cepollaro_harvesting_2025,liu_quantum_2025,esposito2025magic}, even conformal field theories~\cite{hoshino_stabilizer_2025,frau_stabilizer_2025,hoshino_stabilizer_2025-1,Fliss_2021} and Bell's inequalities~\cite{Howard_2013,macedo2025witnessingmagicbellinequalities}, this body of work both analytical and numerical was accelerated by the introduction of a new magic monotone, the \textit{Stabilizer Rényi entropy} (SRE)~\cite{leone2022StabilizerRenyiEntropy}.

Thus, SRE provides a unifying language to quantify how far a quantum state departs from the stabilizer polytope, capturing the essential quantumness required for universal computation. Yet, despite the extensive use of these quantities in both theoretical and experimental settings, much less is known about their statistical properties when quantum states are drawn uniformly from the Hilbert space according to the Haar measure. 

\begin{figure}
    \centering
    \includegraphics[width=0.5\linewidth]{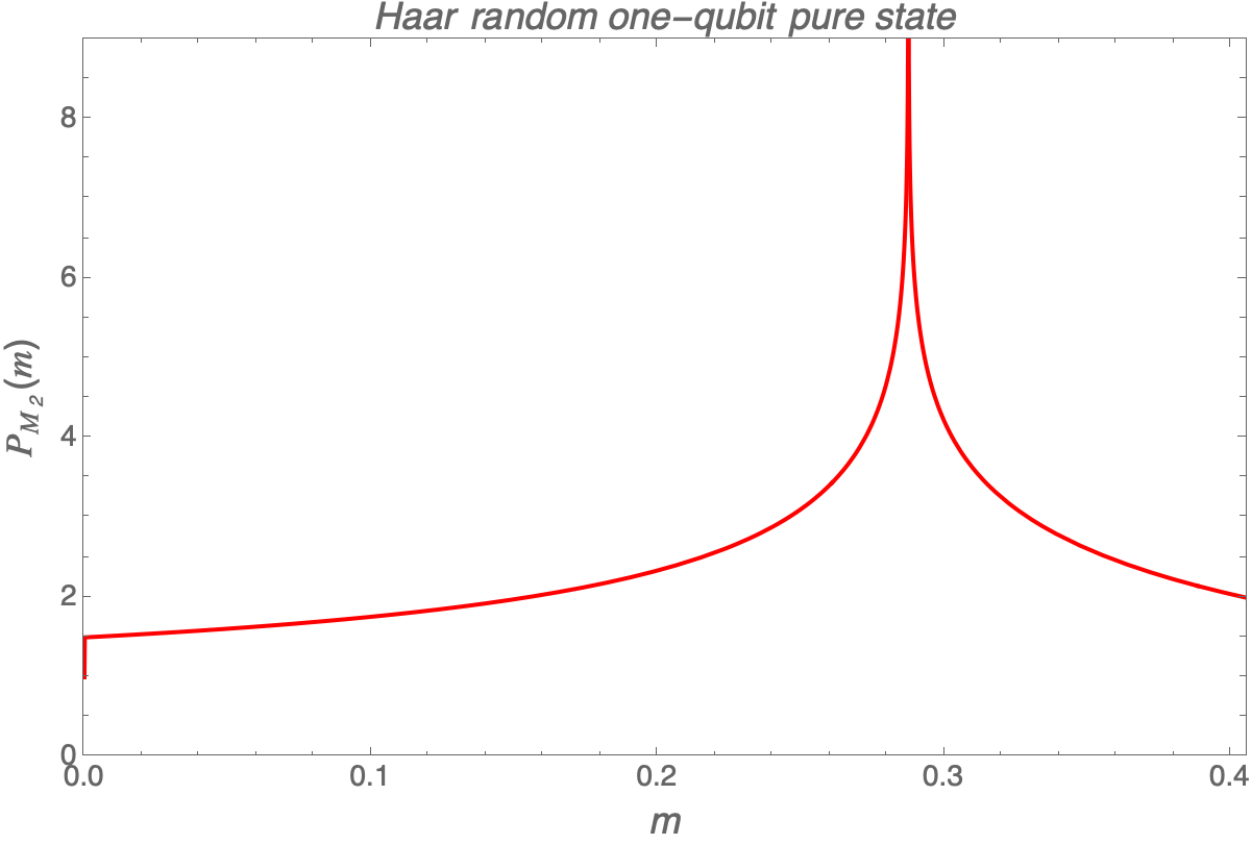}
    \caption{Probability density function, $P_{M_2}(m)$, of the random variable SRE of order two, 
    $m=M_2(\ket{\psi})$
    (see Eq.~\eqref{eq:SRE}), according to the Haar measure for one qubit. The support of $P_{M_2}$ is the interval $[0, \ln (3/2)=0.405\ldots]$. The divergence, of logarithmic type, takes place at $m_c=\ln(4/3)=0.287\ldots$, corresponding to the 12 magic states in the Clifford orbit of $\ket{H}=(\ket{0}+e^{i \pi /4}\ket{1})/\sqrt{2}$. 
    \label{fig:SRE_density}}
\end{figure}

In this work, we study the Haar-induced probability densities of stabilizer purities and their corresponding entropies. The problem of describing these probability density functions (PDFs) in the case of magic is related to the interesting mathematical question of  characterizing the intersection between an $\ell_2$ and an $\ell_{2\alpha}$ sphere. We show that, even for the simplest case of a single qubit, these distributions encode rich geometric information. By mapping the problem to the density of states of a fictitious energy function defined on the Bloch sphere, we uncover the emergence of Van Hove singularities/logarithmic divergencies characteristic of two-dimensional dispersions at saddle points of the energy surface, as already noted in~\cite{557f-6tpb}.
In the case of one qubit we find that the logarithmic divergencies in the PDFs take place at the so-called magic states $\ket{H}$, defined first in the original work of Bravyi and Kitaev~\cite{bravyi2005UniversalQuantumComputation} (see Fig.~\ref{fig:SRE_density}).
In a sense this shows that $\ket{H}$-states are more resilient (in a statistical sense) because there is a large number of magic states with similar value of non-stabilizerness. 
We provide explicit analytical results for the stabilizer entropy and purity with Rényi index $\alpha=2$, numerically validate the predicted logarithmic divergence, and discuss the absence of such singularities in higher-dimensional Hilbert spaces.
Borrowing results for the density of states in solids, one may be led to conclude that van Hove singularities in $d =2$ are present in the probability density of any quantity defined on the Bloch sphere. One subtlety to remember is that the probability densities considered here are defined on a curved base manifold, the (Bloch) sphere, as opposed to a flat manifold, as is the standard case for the density of states in solids. Indeed, we find that other physically interesting quantities defined on the Bloch sphere, such as coherence or mean energy, do not display logarithmic divergence in their PDFs. 
So far, non-stabilizerness as measured by SRE (and likely also by the magic trace distance see the discussion in Sec.~\ref{sec:considerations} \cite{junior2025geometricanalysisstabilizerpolytope}), are the only cases where this logarithmic divergencies are observed.
Finally, we show how stabilizer purities, beside having a clear information-theoretic~\cite{leone2022StabilizerRenyiEntropy} and operational \cite{PhysRevA.107.022429} interpretation, have a direct meaning in terms of fundamental quantum
mechanics concepts, namely \emph{partial incompatibility} of observables and bases.

\section{Measures of non-stabilizerness}

Throughout the paper, we consider $\psi=\ket{\psi}\bra{\psi}$ to
be a pure state on the Hilbert space $\mathcal{H}$ of $L$ qubits
of dimension $d=2^{L}$. Acting on $\mathcal{H}$, there is a natural
and preferential operator basis given by Pauli operators $P\in\mathbb{P}_{L}$,
i.e. $n$-fold tensor products of ordinary Pauli matrices $\1,X,Y,Z$.
The subgroup of unitary matrices that maps Pauli operators to Pauli
operators is called the Clifford group, denoted as $\mathcal{C}_{L}$~\cite{nielsen_chuang_2010,watrous2025understandingquantuminformationcomputation}.
The set of \textit{pure stabilizer states} of $\hi$ is defined as
the orbit of the Clifford group through the computational basis states
$\ket{i}$ (the eigenstates of the $Z$ operator)~\cite{Veitch_2014}: 
\begin{equation}
\PSTAB:=\{C\ket{i}\,,\,C\in\mathcal{C}_{L}\}\,.
\end{equation}
Equivalently, a pure stabilizer state can be defined as the common
eigenstate of $d$ mutually commuting Pauli strings.

Roughly speaking, a quantum circuit that only produces stabilizer
states can be efficiently simulated by a classical computer. More
precisely, the \textit{Gottesman-Knill theorem}~\citep{gottesman1998heisenberg} states that any quantum
process that can be represented by an initial stabilizer state upon
which one performs i) Clifford unitaries; ii) measurements of Pauli
operators; and iii) Clifford operations conditioned on classical randomness;
can be simulated by a classical computer in polynomial time.
Since the set of stabilizer states is by definition closed under Clifford
operations, some resources, such as unitary operations outside the
Clifford group or states not in $\PSTAB$, need to be injected in
the system in order to allow for universal quantum computation. These
non-stabilizer resources, referred to as \textit{non-stabilizerness}
(or \textit{magic}) of the state, have been proven to be a useful
resource for universal quantum computation~\citep{bravyi2005UniversalQuantumComputation}
and several measures have been proposed to quantify them~\citep{howard2017ApplicationResourceTheory,veitch2014ResourceTheoryStabilizer}.
Here we will focus on the essentially only computable measure for
non-stabilizerness, namely the \textit{Stabilizer Rényi Entropy} (SRE)~\citep{leone2022StabilizerRenyiEntropy}:
\begin{definition}
[Stabilizer entropies~\citep{leone2022StabilizerRenyiEntropy}]\label{def:stabilizerentropy}
Let $\ket{\psi}$ be a pure quantum state and $\mathbb{R}\ni\alpha\ge2$.
The $\alpha$-stabilizer R\'{e}nyi entropy (SRE) is defined as 
\begin{equation}
M_{\alpha}(\ket{\psi})\coloneqq\frac{1}{1-\alpha}\ln\Xi_{\alpha}(\ket{\psi})\,,\label{eq:SRE}
\end{equation}
where $\Xi_{\alpha}$'s are the stabilizer purities (SP)\footnote{Stabilizer purities satisfy $\Xi_{2\alpha}^{\frac{\alpha}{\alpha-1}}\le\Xi_{2(\alpha+1)}\le\Xi_{2\alpha}$,
see also~\citep{zhu2016CliffordGroupFails}.}
\begin{equation}
\Xi_{\alpha}(\ket{\psi})\coloneqq\frac{1}{d}\sum_{P\in\mathbb{P}_{L}}|\langle\psi|P|\psi\rangle|^{2\alpha}\,.\label{eq:SRE_SP}
\end{equation}
\end{definition}

Another useful quantifier of non-stabilizerness, 
analogous to the linear entropy of entanglement, is given by the linear
stabilizer entropy defined as~\citep{bittel2025operationalinterpretationstabilizerentropy}: 
\begin{equation}
M_{\alpha}^{\mathrm{lin}}(\psi):=1-\Xi_{\alpha}(\psi)\,.
\end{equation}

Both of these measures are: i) faithful, i.e. $M_{\alpha}(\psi)=M_{\alpha}^{\mathrm{lin}}(\psi)=0\Leftrightarrow\psi\in\PSTAB$;
ii) non-increasing under free operations, that is, operations that send states in $\PSTAB$  to states in $\PSTAB$; and iii) additive under tensor product,
namely $M_{\alpha}(\psi\otimes\sigma)=M_{\alpha}(\psi)+M_{\alpha}(\sigma)$,
or multiplicative for the stabilizer purity $\Xi_{\alpha}$, i.e.
$\Xi_{\alpha}(\psi\otimes\sigma)=\Xi_{\alpha}(\psi)\,\Xi_{\alpha}(\sigma)$.

Both linear and SREs are regarded as good monotones for the
pure-state resource theory of stabilizer computation~\citep{Leone_Bittel_2024}
and have a clear operational meaning~\citep{bittel2025operationalinterpretationstabilizerentropy},
hence numerous schemes have been proposed to study them experimentally~\citep{oliviero2022MeasuringMagicQuantum,Haug_2024,PRXQuantum.4.010301,garcia2025hardness,stratton2025algorithmestimatingalphastabilizerrenyi}.
In recent years, SREs have emerged as a powerful diagnostic of diverse
quantum phenomena. They capture features of error correction~\citep{Niroula_2024}
and measurement-induced phase transitions~\citep{PRXQuantum.5.030332,Fux_2024},
and are closely connected to property testing protocols~\citep{Leone_2024,PhysRevA.107.022429}.
SREs are further related to participation entropy~\citep{PhysRevA.108.042408},
which provides insight into Anderson localization~\citep{castellani2005multifractal}
and many-body localization~\citep{PhysRevB.80.184421}. Beyond these
structural aspects, SREs also exhibit a fruitful interplay with non-stabilizerness
and out-of-time-order correlators (OTOCs)~\citep{Leone_2021,Garcia_2023}.
Higher-order extensions reveal additional links between OTOCs, nonstabilizerness,
quantum chaos~\citep{Leone_2021}, and state certification~\citep{PhysRevA.107.022429},
further underlining the versatility of SREs as a lens on quantum dynamics.

\section{Probability densities of measures of non-stabilizerness}

Due to normalization and invariance under phase change, the set of
pure states on $\mathcal{H}$ can be identified with $\mathbb{C}P^{d-1}$,
the complex projective space of dimension $d-1$~\citep{nielsen_chuang_2010}.
On this manifold, of real dimension $2d-2$, there is a unique, unitarily
invariant measure $d\psi$~\cite{Bengtsson_Zyczkowski_2006,wootters1990random}. This measure is induced by the Haar measure
$dU$ on the corresponding unitary group $U(d)$ when applied to a
fiducial state $|\psi_{0}\rangle\in\mathcal{H}$, i.e. 
\begin{equation}
\int_{\mathbb{C}P^{d-1}}d\psi\, f(|\psi\rangle)=\int_{U(d)}dUf(U|\psi_{0}\rangle)\,,\label{eq:Haar0}
\end{equation}
for any integrable function $f$. A function from the set of pure
states to $\mathbb{R}$ becomes a real random variable when $\mathbb{C}P^{d-1}$
is equipped with this uniform measure. Expectation values of $f$
are computed with Eq.~\eqref{eq:Haar0} via $\mathbb{E}_{\psi}[f(|\psi\rangle)]=\mathbb{E}_{U}[f(U|\psi_{0}\rangle)]=\int dUf(U|\psi_{0}\rangle)$~\cite{Mele_2024,collins2006integration,watrous2018theory,Collins_2015}.
In this work, we are interested in the probability density function
(PDF) of the SRE when $|\psi\rangle$ is distributed uniformly over
the set of pure states, in the same spirit of entanglement~\cite{giraud2007purity}. In particular we will concentrate on the single
qubit case, whence $d=2$, although we will make some comments for
larger $d$. Since $M_{\alpha}$ is a simple function of $\Xi_{\alpha}$,
the PDF of $M_{\alpha}$ can be obtained from that of $\Xi_{\alpha}$
with a change of variable. Specifically,
\begin{equation}
P_{M_{\alpha}}(m)=|1-\alpha|e^{(1-\alpha)m}P_{\Xi_{\alpha}}(e^{(1-\alpha)m})\,.
\end{equation}
Hence, we are led to compute $P_{\Xi_{\alpha}}(\xi)$ which can be
formally written as
\begin{equation}
P_{\Xi_{\alpha}}(\xi):=\mathbb{E}_{\psi}[\delta\left(\Xi_{\alpha}(\ket{\psi}-\xi\right)]=\int_{\mathbb{C}P^{d-1}}\!\!d\psi\,\delta\left(\Xi_{\alpha}(\ket{\psi})-\xi\right)\,.
\end{equation}
In the following sections
we are going to derive some of its features as well as exact results.

\section{Van Hove singularities for general \texorpdfstring{$\alpha$}{TEXT}}

\begin{figure}
\begin{centering}
\includegraphics[width=7cm]{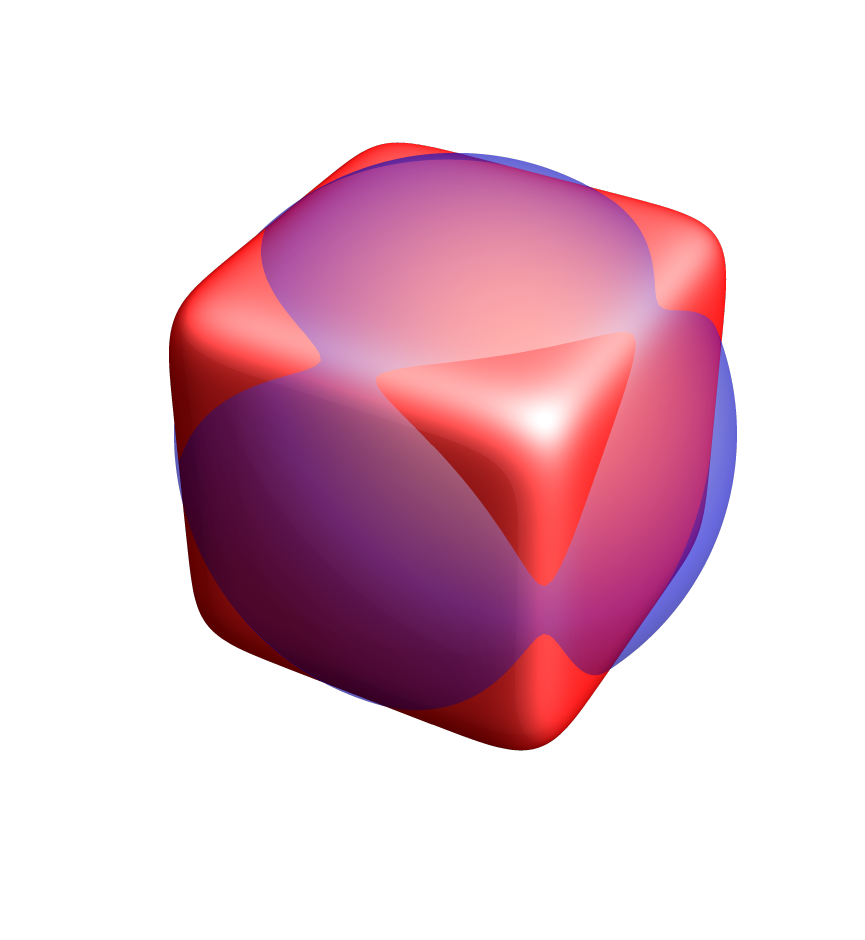}\includegraphics[width=7cm]{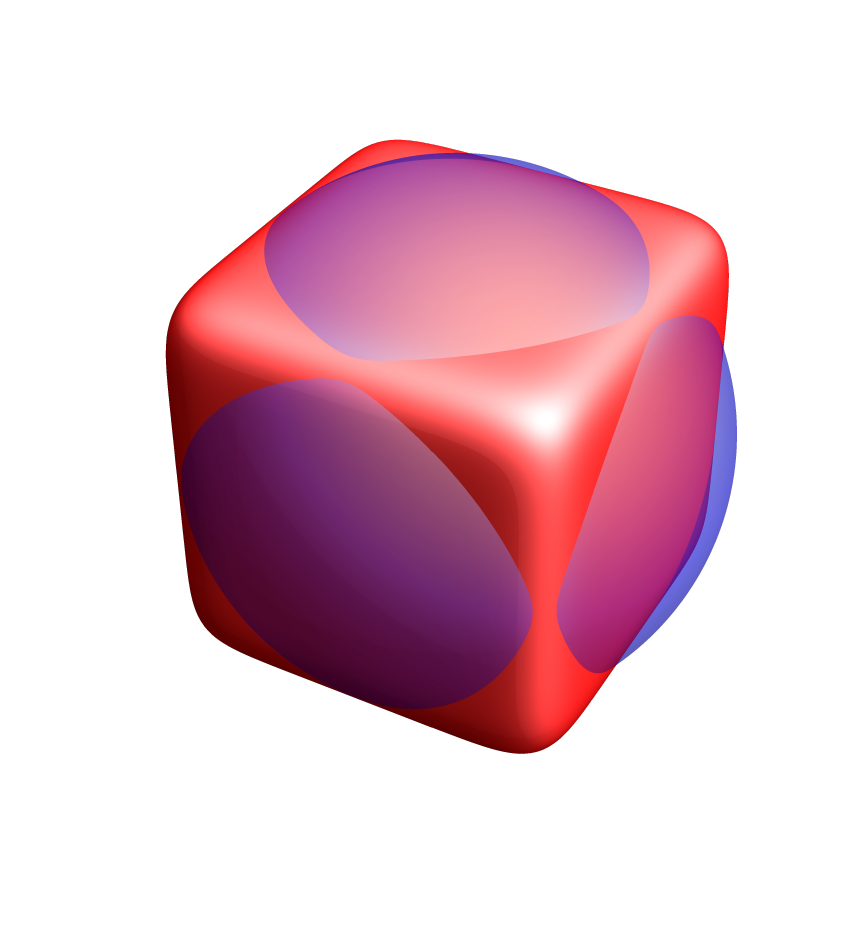}
\par\end{centering}
\caption{Intersection between $\ell^{2\alpha}$ and $\ell^{2}$spheres, here
for $\alpha=4$. Left panel $n=0.12<n_{c}$, right
panel $n=0.13>n_{c}$. \label{fig:intersection_3D}}
\end{figure}

From here on we focus on the single qubit case. A single qubit pure
state can be written as 
\begin{equation}
\psi=\frac{\1+\boldsymbol{n}\cdot\boldsymbol{\sigma}}{2}\,,
\label{eq:qubit}
\end{equation}
where $\bm{n}=(n_{1},n_{2},n_{3})$ belongs to the Bloch's sphere
as defined by the $\ell^{2}$-norm, i.e.~$\norm{\bm{n}}_{2}=1$,
while $\bm{\sigma}=(X,Y,Z)$ is a vector
of Pauli matrices. The stabilizer purities in Eq.~\eqref{eq:SRE_SP} become 
\begin{align}
\Xi_{\alpha}(\ket{\psi}) & =\frac{1}{2}\sum_{P\in\mathbb{P}_{1}}|\langle\psi|P|\psi\rangle|^{2\alpha}=\frac{1+\norm{\bm{n}}_{2\alpha}^{2\alpha}}{2}\\
 & =:\frac{1+N_{\alpha}}{2}\in[\frac{1+3^{1-\alpha}}{2},1]\,,\label{eq:Xi_2a}
\end{align}
where we defined $N_{\alpha}=\norm{\bm{n}}_{2\alpha}^{2\alpha}=\sum_{j=1}^{3}|n_{j}|^{2\alpha}$.
The maximum value of $N_{\alpha}$ is one, attained when ${\bf {n}}$
is $\pm$ one of the three normalized coordinate vectors $\hat{{\bf {x}}},\hat{{\bf {y}}},\hat{{\bf {z}}}$,
while the minimum of $N_{\alpha}$ takes place when $|n_{1}|=|n_{2}|=|n_{3}|=1/\sqrt{3}$
for which $N_{\alpha}=3^{1-\alpha}$. This explains the bound in Eq.~(\ref{eq:Xi_2a}).
We focus our analysis on the quantity $N_{\alpha}=\norm{\bm{n}}_{2\alpha}^{2\alpha}$
and then turn back to the study of Eqns.~(\ref{eq:SRE}-\ref{eq:SRE_SP})
through a change of variable, i.e., via 
\begin{equation}
P_{\Xi_{\alpha}}(\xi)=2P_{N_{\alpha}}(2\xi-1)\,,\quad P_{M_{\alpha}}(m)=2(\alpha-1)e^{(1-\alpha)m}P_{N_{\alpha}}(2e^{(1-\alpha)m}-1)\,.\label{eq:change_variable}
\end{equation}

We use spherical coordinates to enforce the constraint $\norm{\bm{n}}_{2}=1$,
hence, using the coarea formula, the PDF of $N_{\alpha}(\vartheta,\phi)=\left|\sin\left(\vartheta\right)\right|^{2\alpha}\left(\left|\cos\left(\phi\right)\right|^{2\alpha}+\left|\sin\left(\phi\right)\right|^{2\alpha}\right)+\left|\cos\left(\vartheta\right)\right|^{2\alpha}$
can be written as 
\begin{align}
P_{N_{\alpha}}(n) & =\int_{0}^{2\pi}\int_{0}^{\pi}\frac{\sin\vartheta d\vartheta d\phi}{4\pi}\delta(N_{\alpha}(\vartheta,\phi)-n)\nonumber \\
 & =\int_{N_{\alpha}^{-1}(n)}\frac{d\sigma_{\lambda}}{4\pi}\frac{1}{\norm{\nabla N_{\alpha}(\vartheta(\lambda),\phi(\lambda))}_{2}}\,,\label{eq:PN_coarea}
\end{align}
with $d\sigma_{\lambda}=\sqrt{\left(\frac{d\vartheta}{d\lambda}\right)^{2}+\sin^{2}\vartheta\left(\frac{d\phi}{d\lambda}\right)^{2}}d\lambda$
the induced surface measure on the level set, a one-dimensional set
in this case.

At this point we would like to remark that Eq.~(\ref{eq:PN_coarea})
for $P_{N_{\alpha}}(n)$ is exactly the density of states (DOS) of
a system with energy dispersion given by $N_{\alpha}(\vartheta,\phi)$, for which the coordinates
$\boldsymbol{\vartheta}:=(\vartheta,\phi)$ play the role of momenta and 
whose first Brillouin zone is the sphere.
From this observation, we can already draw some conclusions from the
vast existing literature regarding the behavior of the density of
states in solids. Eq.~(\ref{eq:PN_coarea}) suggests that one may
have divergencies of the DOS, called Van Hove singularities, at locations
where the gradient $\nabla N_{\alpha}$ is zero (the dispersion is
flat). Such critical points correspond to local maxima or minima of
$N_{\alpha}$ or saddle points. Generally, in two dimensions, as is
the dimension of the sphere, locations of maxima or minima of the
dispersions do not give rise to a divergent DOS, but rather simply
step-wise singularities corresponding to the behavior at the edge.
Instead saddle points give rise to logarithmic singularities in the
DOS~\citep{yuan_classification_2020}. In the following we will carefully
prove that these predictions are indeed correct. 

\begin{figure}
\begin{centering}
\includegraphics[width=7cm]{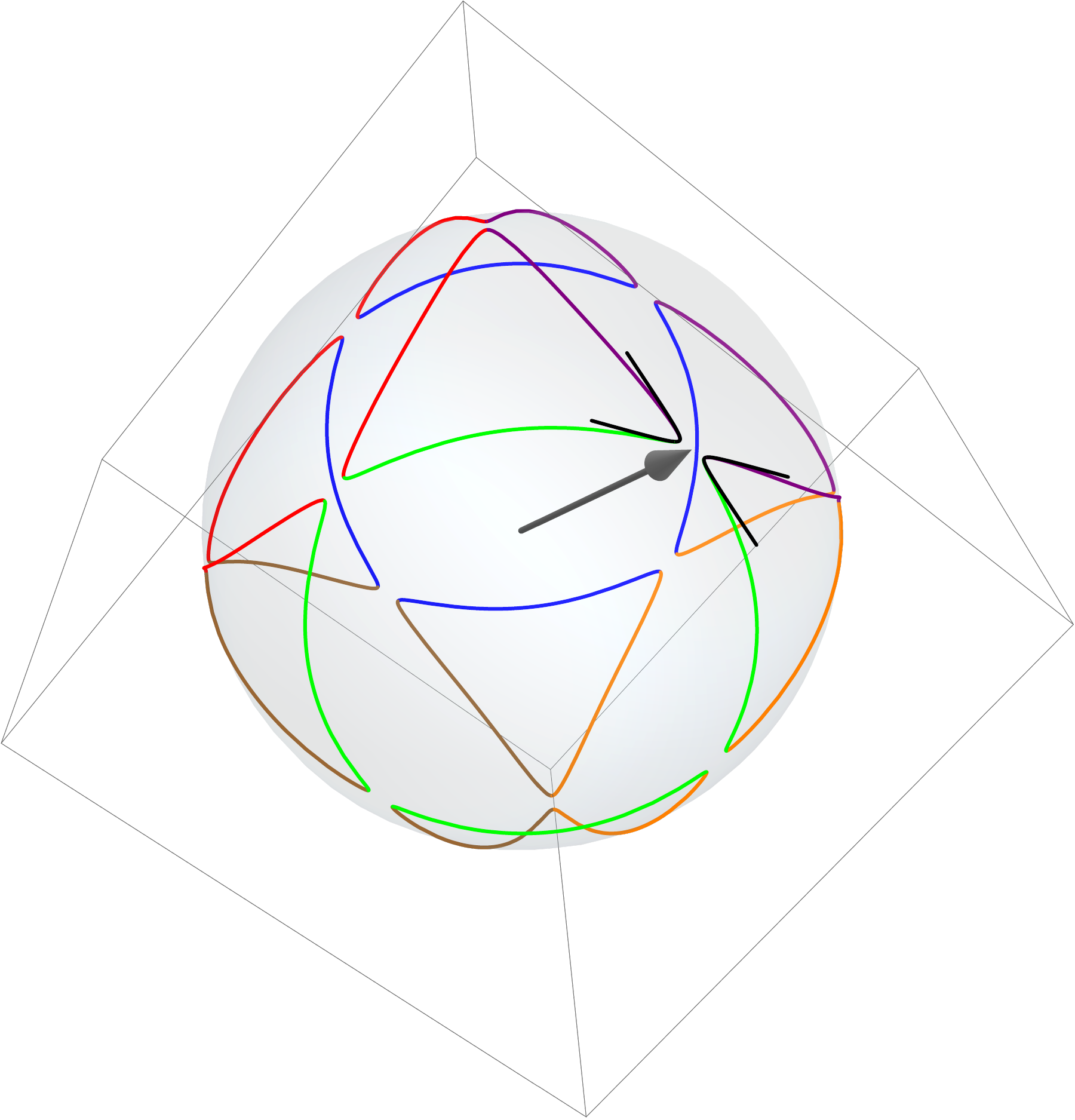}\includegraphics[width=7cm]{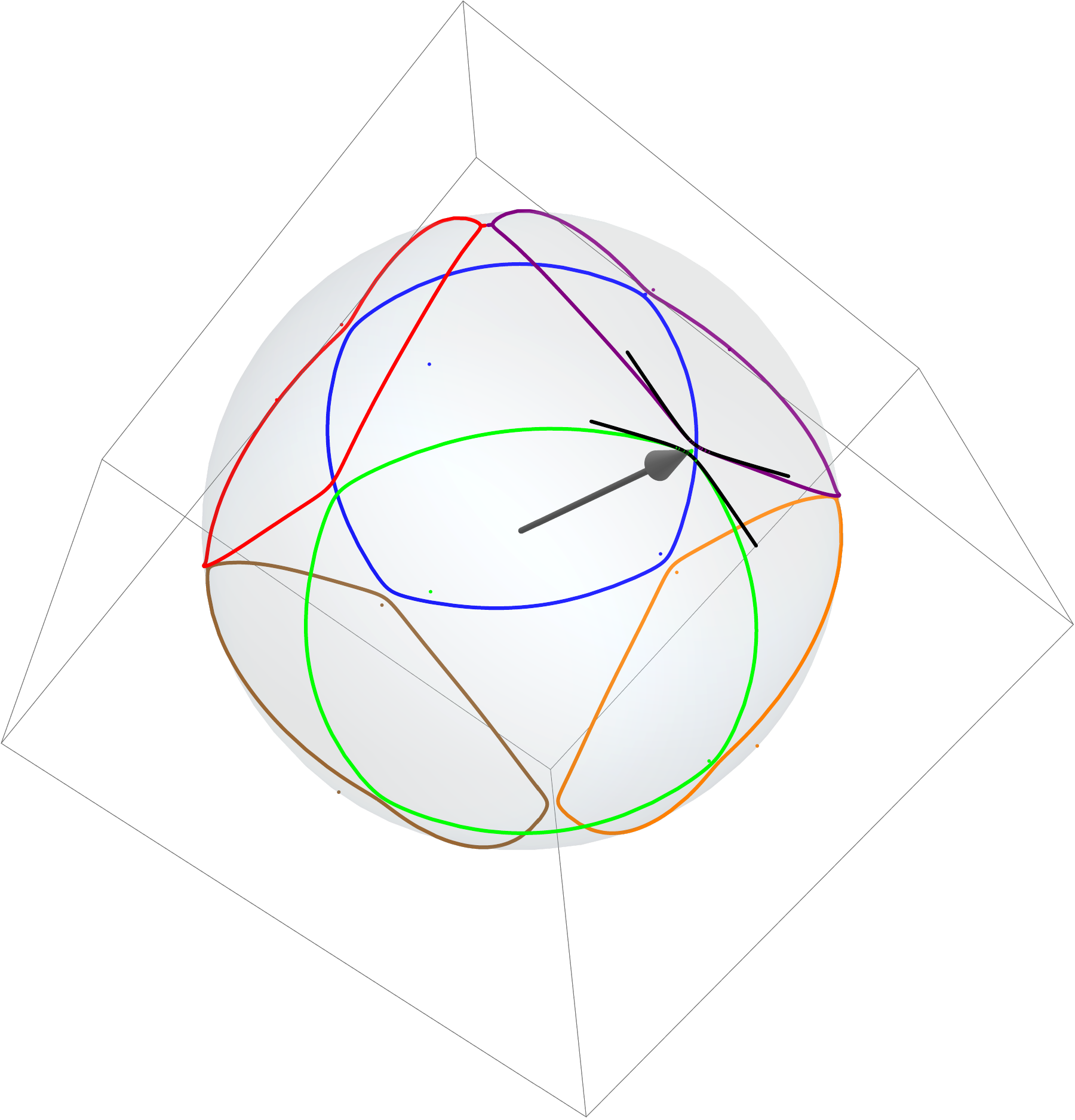}
\par\end{centering}
\caption{Integration regions to compute the stabilizer purity PDF for $\alpha=4$.
The grey arrow is the critical point $\boldsymbol{n}_{c}=\left(1/\sqrt{2},1/\sqrt{2},0\right)$
and the black curves are the hyperbolae where integration takes place
for $n$ close to $n_{c}=1/8$. Left panel $n=0.124<n_{c}$, right
panel $n=0.126>n_{c}$. \label{fig:intersection_loops}}
\end{figure}

The integration region in Eq.~(\ref{eq:PN_coarea}) is the intersection
between the $\ell^{2\alpha}$ sphere of radius $n^{1/2\alpha}$,  i.e. $\left|n_{1}\right|^{2\alpha}+\left|n_{2}\right|^{2\alpha}+\left|n_{3}\right|^{2\alpha}=n$,
and the $\ell^{2}$ sphere of radius one, i.e.~$n_{1}^{2}+n_{2}^{2}+n_{3}^{2}=1$.
Since $\alpha>1$ the $\ell^{2\alpha}$ sphere resembles a cube with rounded
edges, see Figure \ref{fig:intersection_3D}. When $n$ is close to one,
the intersection gives rise to eight closed loops at the vertices
of this rounded cube. Instead, when $n$ is close to the lower bound $3^{1-\alpha}$,
the intersection gives rise to closed loops around the faces of the
smoothed cube so that in this case there are six closed loops, see
Figure \ref{fig:intersection_loops}. As $n$ varies smoothly from the
lower bound to the upper bound, there must be a critical value $n_{c}$
where the six loops and the eight loops coexist. It is natural to
suppose that for this value of $n$, the gradient of $N_{\alpha}$
vanishes at some point. Notice that \emph{critical points} of the
function $N_{\alpha}=\|\bm{n}\|_{2\alpha}^{2\alpha}$ under the constraint
$\|\bm{n}\|_{2}^{2}=1$, that is, points where the gradient is zero,
correspond to \emph{isolated points} on the intersection of the unit
sphere and the $\ell_{2\alpha}$-sphere defined by $N_{\alpha}=n$.
The full set of solutions of the system 
\begin{equation}
\|\bm{n}\|_{2}^{2}=1,\quad\text{and}\quad\|\bm{n}\|_{2\alpha}^{2\alpha}=n\,,
\end{equation}
is generally a \emph{1-dimensional manifold} (a curve) embedded in
the sphere, as it represents the intersection of two smooth surfaces
in $\mathbb{R}^{3}$. Therefore, the solutions form continuous, smooth
curves, except at critical points where these curves can change topology
or number of connected components, as shown in Fig.~\ref{fig:intersection_loops}.
In spherical coordinates, using the substitution $s=\sin\left(\vartheta\right)^{2}$,
the equation $N_{\alpha}(\vartheta,\phi)=n$ reduces to a polynomial
of degree $\alpha$ in $s$. Therefore, in principle, solutions are
available in terms of radicals (after a careful examination of various
positivity conditions) for $\alpha=2,3,4$, but the resulting expressions
for the integral in Eq.~(\ref{eq:PN_coarea}) are kilometric and
unpractical for $\alpha>2$. We will give an explicit integral representation
for $P_{N_{\alpha}}(n)$ for $\alpha=2$ in Section \ref{sec:Case_a2}
using a different method. 

We first find the critical points of the function $N_{\alpha}(\boldsymbol{n})$
where $\boldsymbol{n}\in\mathbb{S}^{2}=\left\{ \boldsymbol{n}|\left\Vert \boldsymbol{n}\right\Vert _{2}=1\right\} $,
i.e.~points on the sphere where $\nabla N_{\alpha}=0$. 
In order to do so, we enforce the constraint via Lagrange multipliers
and define the following Lagrangian:
\begin{equation}
\mathcal{L}(n_{1},n_{2},n_{3},\lambda)=\sum_{i=1}^{3}|n_{i}|^{2\alpha}-\lambda\left(\sum_{i=1}^{3}n_{i}^{2}-1\right)\,.
\end{equation}
Setting the partial derivatives to zero we get 
\begin{equation}
\frac{\partial\mathcal{L}}{\partial n_{i}}=2\alpha\,\text{sign}(n_{i})\,|n_{i}|^{2\alpha-1}-2\lambda n_{i}=0\,.
\end{equation}
The solution of the above equation is either $n_{i}=0$, or, assuming $n_{i}\ne0$,
\begin{equation}
\alpha|n_{i}|^{2\alpha-2}=\lambda\,,
\end{equation}
which implies that all nonzero components of $\bm{n}$ have the same
absolute value.The function $N_{\alpha}:\mathbb{S}^{2}\to\mathbb{R}$
is invariant under the action of the \emph{hyperoctahedral group}
$G=S_{3}\ltimes\mathbb{Z}_{2}^{3}$ \cite{Heinrich_2019}, which consists of all permutations
and sign changes of the components of $\bm{n}$. By Palais principle~\citep{palais1979principle},
critical points of $N_{\alpha}$ must occur in $G$-symmetric configurations,
i.e., vectors that are invariant under some subgroup of $G$. Up to
symmetry, the critical points of $N_{\alpha}$ on $\mathbb{S}^{2}$
fall into three classes:

\begin{equation}
\begin{split}
C_1&= \{(\pm1,0,0)\,, (0,\pm1 ,0)\,, (0,0,\pm1)\},\quad N_{\alpha}=1\\
C_2 &= \{\left(\pm\frac{1}{\sqrt{2}},\pm\frac{1}{\sqrt{2}},0\right)\,, \left(\pm\frac{1}{\sqrt{2}},0,\pm\frac{1}{\sqrt{2}}\right)\,, \left(0,\pm\frac{1}{\sqrt{2}},\pm\frac{1}{\sqrt{2}}\right)\},\quad N_{\alpha}=2^{1-\alpha}\\
C_3& =\{\left(\pm\frac{1}{\sqrt{3}},\pm\frac{1}{\sqrt{3}},\pm\frac{1}{\sqrt{3}}\right)\},\quad N_{\alpha}=3^{1-\alpha}
\end{split}\,.\label{eq:classes}
\end{equation}
There are respectively 6, 12 and 8 points in each class. The first,
respectively third, class correspond to absolute maximum, respectively
minimum, of $N_{\alpha}$, namely stabilizer states and maximal magic states ( $\ket{T}$-type states in \cite{bravyi2005UniversalQuantumComputation}). The corresponding values of $N_\alpha$, $1,3^{1-\alpha}$,
are the edge of the support of $P_{N_{\alpha}}(n)$, $[3^{1-\alpha},1]$,
and for what we have said previously $P_{N_{\alpha}}(n)$ has a
simple stepwise singularity around these points where it raises from
zero to a finite value at the lower edge or goes to zero from a finite
value at the upper edge. 
The 12 points in the second class, taking
place at $n=n_{c}:=2^{1-\alpha}$, are precisely the locations where
the 8 loops for $n<n_{c}$ touch the six loops for $n>n_{c}$.
The corresponding states are precisely the so called $\ket{H}$-states defined first in \cite{bravyi2005UniversalQuantumComputation}, given by the Clifford orbit of $\ket{H}\bra{H}=\frac{1}{2}\left(\id+\frac{X+Z}{\sqrt{2}}\right)$. 
As
we will show below, these are saddle points of $N_{\alpha}$ and give
rise to logarithmic, Van Hove singularities in $P_{N_{\alpha}}(n)$.
Notice that the position of the singularity is given by
\begin{equation}
    \Xi_{\alpha}|_c= \frac{1+2^{1-\alpha}}{2}, \quad  M_\alpha|_c = \frac{1}{1-\alpha} \ln \left( \frac{1+2^{1-\alpha}}{2} \right).
\end{equation}

In order to prove this claim we need to find, to leading order, the
curves resulting from the intersection of $N_{\alpha}=n$ and $\norm{\bm{n}}_{2}^{2}=1$
when $n$ is close to $n_{c}$, for generic $\alpha\geq2$. We fix
$\boldsymbol{n}_{c}=\left(1/\sqrt{2},1/\sqrt{2},0\right)$, so in
spherical coordinate we set $\vartheta=\pi/2+\vartheta'$ and $\phi=\pi/4+\phi'$,
where $\vartheta'=\phi'=0$ corresponds to $\boldsymbol{n}_{c}$.
Since $\alpha>1$ we obtain, up to leading order 
\begin{equation}
\begin{split}\norm{\bm{n}}_{2\alpha}^{2\alpha} & \simeq2^{1-\alpha}+2^{2-\alpha}\alpha(\alpha-1)\phi'{}^{2}-2^{1-\alpha}\alpha\vartheta'{}^{2}\\
 & =n_{c}+2^{2-\alpha}\alpha(\alpha-1)\phi'{}^{2}-2^{1-\alpha}\alpha\vartheta'{}^{2}\,,
\end{split}
\end{equation}
explicitly showing that $\boldsymbol{n}_{c}$ is a saddle point (the
linear terms are zero and the Hessian has a positive and a negative
eigenvalue). Equating the above to $n$ we see that the constraint
becomes the equation of an hyperbola in the variables $\phi',\vartheta'$:
\begin{equation}
2^{2-\alpha}\alpha(\alpha-1)\phi'{}^{2}-2^{1-\alpha}\alpha\vartheta'{}^{2}=n-n_{c}.
\end{equation}
For $n>n_{c}$ we can write 
\begin{equation}
\phi'=\pm\sqrt{\frac{n-n_{c}}{2^{2-\alpha}\alpha(\alpha-1)}+\frac{1}{2(\alpha-1)}\vartheta'{}^{2}}\,.\label{eq:constraint}
\end{equation}
Exactly at $n=n_{c}$ the hyperbolae degenerate
into two straight lines (the asymptotes for $n\neq n_{c}$)
\begin{equation}
\phi'=\pm\sqrt{\frac{1}{2(\alpha-1)}}\vartheta',
\end{equation}
however on these lines the integrand is infinite so we need to keep
$n\neq n_{c}$. We now expand the gradient around the same point and
get 
\begin{equation}
\norm{\nabla N_{\alpha}}_{2}^{2}\simeq4^{3-\alpha}(\alpha-1)^{2}\alpha^{2}\phi'{}^{2}+4^{2-\alpha}\alpha^{2}\vartheta'{}^{2}\,,
\end{equation}
which, after imposing the constraint Eq.~(\ref{eq:constraint}),
becomes 
\begin{equation}
\norm{\nabla N_{\alpha}}_{2}^{2}\simeq2^{\alpha}4^{2-\alpha}\alpha(\alpha-1)(n-n_{c})+4^{2-\alpha}\alpha^{2}(2\alpha-1)\vartheta'{}^{2}\,.
\end{equation}
Plugging in the line element and expanding to leading order, we obtain
the most diverging contribution to the integral Eq.~(\ref{eq:PN_coarea})
for $n\to n_{c}$ 
\begin{equation}
\int_{-\delta}^{\delta}d\vartheta'\frac{1}{\sqrt{2^{\alpha}4^{2-\alpha}\alpha(\alpha-1)(n-n_{c})+4^{2-\alpha}\alpha^{2}(2\alpha-1)\vartheta'{}^{2}}}=-\frac{\ln\left(n-n_{c}\right)}{2^{2-\alpha}\alpha\sqrt{2\alpha-1}}+O(1)\,,
\end{equation}
where $\delta$ is any small positive constant and we used the following
identity valid for $\epsilon\to0$ (and $a,b>0$), \citep{gradshteyn1988tables}
\begin{equation}
\int_{-\delta}^{\delta}\frac{dy}{\sqrt{a\epsilon+by^{2}}}=-\frac{\ln\left(\epsilon\right)}{\sqrt{b}}+O(1)\,.\label{eq:divergence_identity}
\end{equation}
Overall we obtain, for $n\to n_{c}^{+}$
\begin{equation}
P_{N_{\alpha}}(n)\propto-\ln\left(n-n_{c}\right).
\end{equation}
For $n<n_{c}$ the roles of $\phi'$ and $\vartheta$'
are reversed. Proceeding analogously we obtain, for $n\to n_{c}$,
\begin{equation}
P_{N_{\alpha}}(n)\propto-\ln\left|n-n_{c}\right|. \label{eq:log_divergence}    
\end{equation}
That is, the anticipated logarithmic divergence of the PDF when $n\to n_{c}$.
Similar expansions show that, close to the minima or maxima of $N_{\alpha}$,
the integrand in Eq.~(\ref{eq:PN_coarea}) has only square root singularities,
which are integrable and give rise to a finite $P_{N_{\alpha}}(n)$. 

The same logarithmic divergence can be observed for the SRE Eq.~\eqref{eq:SRE}, and the SP Eq.~\eqref{eq:SRE_SP}, using Eq.~\eqref{eq:change_variable} (see Appendix \ref{App:Change_variables})
\begin{equation}
    P_{\Xi_\alpha}(\xi) \propto -\ln \left|\xi-\frac{1+n_{c}}{2}\right|\, \quad P_{M_\alpha}(m) \propto - \ln \abs{m- \frac{1}{1-\alpha} \ln \left( \frac{1+n_c}{2} \right)} \,.
    \label{eq:SRE_SP_expantion_nc}
\end{equation}
\section{Case \texorpdfstring{$\alpha=2$}{TEXT} }

\label{sec:Case_a2}

The case $\alpha=2$ can be carried out in greater detail, and it is
also the most important as it corresponds to the simplest non-stabilizer monotone.
We obtain the PDF of $N_{2}=\left\Vert \boldsymbol{n}\right\Vert _{4}^{4}$
by first computing its characteristic function and then Fourier transforming.
First, going to spherical coordinates, we write 
\begin{equation}
\left\Vert \boldsymbol{n}\right\Vert _{4}^{4}=\sin(\vartheta)^{4}+\cos\left(\vartheta\right)^{4}-\frac{\sin\left(2\phi\right)^{2}}{2}\sin(\vartheta)^{4}\,,
\end{equation}
so that the characteristic function of $N_{2}$ reads 
\begin{align}
\chi_{N_{2}}\left(k\right) & =\mathsf{E}\left[e^{ikN_{2}}\right]\\
 & =\int_{0}^{2\pi} \int_{0}^{\pi}\frac{d\phi\, d\vartheta\sin\left(\vartheta\right)}{4\pi}\,\exp\left[-ik\left(\sin\left(2\phi\right)^{2}/2\right)\sin\left(\vartheta\right)^{4}+ik\cos\left(\vartheta\right)^{4}+ik\sin\left(\vartheta\right)^{4}\right].
\end{align}
Now we use 
\begin{equation}
\int_{0}^{2\pi}\frac{d\phi}{2\pi}e^{-iy\sin\left(2\phi\right)^{2}}=e^{-iy/2}J_{0}\left(y/2\right),
\end{equation}
where $J_{0}$ is the Bessel function of the first kind and $y=k\sin\left(\vartheta\right)^{4}/2$,
whence we obtain
\begin{equation}
\chi_{N_{2}}\left(k\right)=\frac{1}{2}\int_{0}^{\pi}d\vartheta\sin\left(\vartheta\right)\exp\left[ik\cos\left(\vartheta\right)^{4}+ik\frac{3}{4}\sin\left(\vartheta\right)^{4}\right]J_{0}\left(\frac{k\sin\left(\vartheta\right)^{4}}{4}\right).
\end{equation}
We now perform a Fourier transform to obtain the PDF. We use the following identity,
with $\theta$ the Heaviside function~\citep{gradshteyn1988tables}
\begin{equation}
\int_{\mathbb{R}}e^{ikx}J_{0}(ax)dx=2\frac{\theta\left(1-\left|k/a\right|\right)}{\sqrt{a^{2}-k^{2}}}=2\frac{\theta\left(a-\left|k\right|\right)}{\sqrt{a^{2}-k^{2}}},\quad\mathrm{for}\,a>0\,,
\end{equation}
with $x=\cos\left(\vartheta\right)$ and 
\begin{align}
a(x) & =\left(1-x^{2}\right)^{2}\\
k(x) & =3\left(1-x^{2}\right)^{2}+4x^{4}-4n\,.
\end{align}
\begin{figure}
\centering \includegraphics[width=0.7\linewidth]{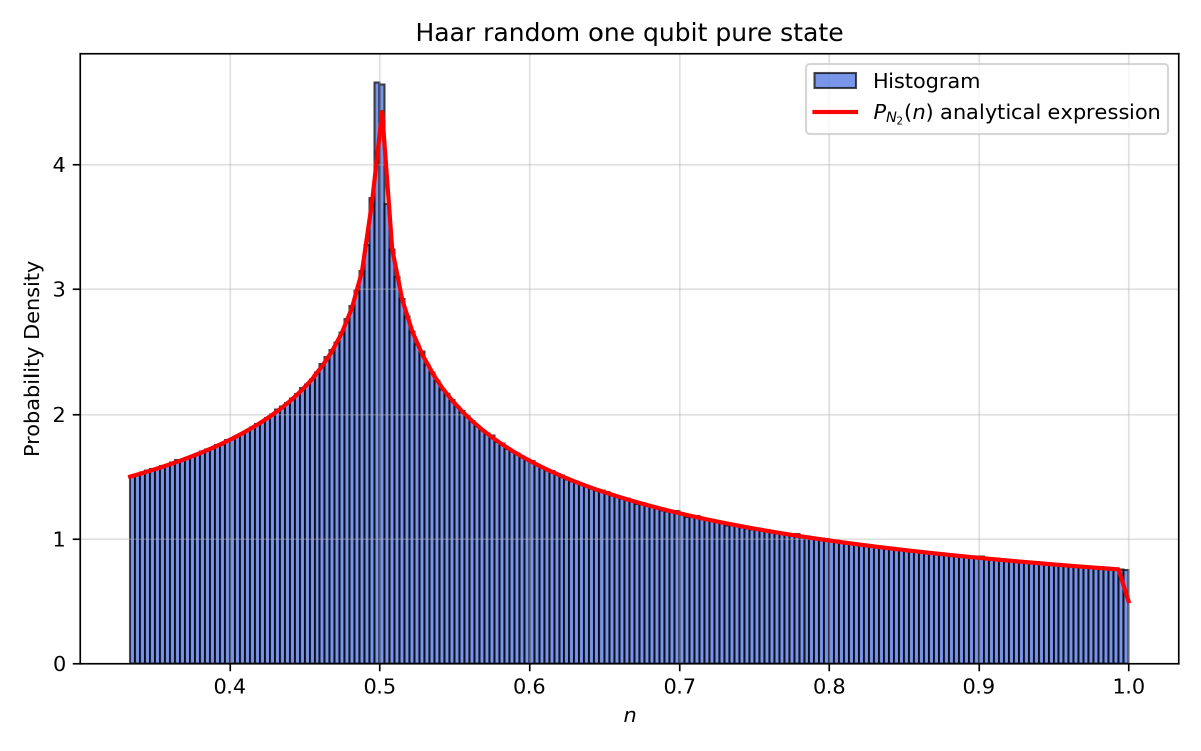}
\caption{In blue the numerical probability density function of $P_{N_2}(\textit{n})$ using $N_{sample}=10^{7}$ one-qubit
random pure states extracted according to the Haar measure. In red the theoretical analytical distribution Eq.~\eqref{eq:PN2}.}
\label{fig:PDF_1_histogram}
\end{figure}
Hence, we obtain the final form of the PDF using the parity of the integrand
\begin{equation}
P_{N_{2}}(n)=\frac{4}{\pi}\int_{0}^{1}dx\frac{\theta\left(\left(1-x^{2}\right)^{2}-\left|3\left(1-x^{2}\right)^{2}+4x^{4}-4n\right|\right)}{\sqrt{\left(1-x^{2}\right)^{4}-\left(3\left(1-x^{2}\right)^{2}+4x^{4}-4n\right)^{2}}}\,.
\end{equation}
For $n\in[1/3,1/2)$ the $\theta$ constraints the integration in $x\in\left[x_{-},x_{+}\right]$,
where $x_{\pm}$ are the roots in $[0,1]$ of $a(x)=k(x)$ which are
\begin{equation}
x_{\pm}=\left(\frac{1\pm\sqrt{6n-2}}{3}\right)^{1/2}.
\end{equation}
For $n\in(1/2,1]$ the integration region $\left|k(x)\right|<a(x)$
becomes $x\in[0,y_{-})$ and $x\in(y_{+},x_{+})$. Where $y_{\pm}$
are the roots of $a(x)=-k(x)$ in $[0,1]$. These are
\begin{equation}
y_{\pm}=\left(\frac{1\pm\sqrt{2n-1}}{2}\right)^{1/2}.
\end{equation}
All in all we obtain (see figure \ref{fig:PDF_1_histogram})
\begin{equation}
P_{N_{2}}(n)=\begin{cases}
\frac{4}{\pi}\int_{x_{-}}^{x_{+}}dx\frac{1}{\sqrt{\left(1-x^{2}\right)^{4}-\left(3\left(1-x^{2}\right)^{2}+4x^{4}-4n\right)^{2}}} & \mathrm{for}\,n\in[1/3,1/2)\\
\frac{4}{\pi}\left(\int_{0}^{y_{-}}dx+\int_{y_{+}}^{x_{+}}dx\right)\frac{1}{\sqrt{\left(1-x^{2}\right)^{4}-\left(3\left(1-x^{2}\right)^{2}+4x^{4}-4n\right)^{2}}} & \mathrm{for}\,n\in(1/2,1]
\end{cases}\,\,.\label{eq:PN2}
\end{equation}
Moreover $P_{N_{2}}(n)=0$ for $n<1/3$ and $n>1$ and is infinite
for $n=1/2$. The PDF of the two-SRE (and of the two-stabilizer purity)
can be obtained via Eq.~(\ref{eq:change_variable}) and the result is shown in Fig.~\ref{fig:SRE_density}.

In particular,
using Eq.~(\ref{eq:PN2}), we obtain the exact mean value of the SRE for
one qubit: 
\begin{equation}
\mathbb{E}_{\psi}[M_{2}(\psi)]=\int_{0}^{1}dx\ln\left(\frac{16}{7x^{4}-6x^{2}+4\sqrt{3x^{8}-5x^{6}+8x^{4}-5x^{2}+3}+7}\right)\simeq0.228921\,,
\end{equation}
see Appendix \ref{App:Mean_SRE_1qubit} for more details.

In the remainder of this chapter we explicitly check the form of the
divergence of $P_{N_{2}}(n)$ for $n\to n_{c}$ using the explicit
expression (\ref{eq:PN2}). Note that the argument of the square root
can be written as 
\begin{equation}
-48\left(x^{2}-x_{-}^{2}\right)\left(x^{2}-x_{+}^{2}\right)\left(x^{2}-y_{-}^{2}\right)\left(x^{2}-y_{+}^{2}\right),
\end{equation}
hence when $x$ is close to the border of integration, $x_{\pm}$
the integrand behaves as 
\begin{equation}
\frac{1}{\sqrt{\left|x-x_{\pm}\right|}},
\end{equation}
which is integrable and leads to a bounded $P_{N_{2}}(n)$. However, when $n\to1/2^{-}$ the integrand develops
another singularity. In this case $y_{\pm}^{2}\to1/2$, and around
$x=\left.y_{-}\right|_{n=1/2}=1/\sqrt{2}$, the integrand
behaves as 
\begin{equation}
\frac{1}{\left|x-\frac{1}{\sqrt{2}}\right|},
\end{equation}
which is not integrable and leads to a divergence of $P_{N_{2}}(n)$.
The same happens for $n\to1/2^{+}$. We can guess that the
divergence will be of logarithmic type, however let us proceed with
order. When $y_{\pm}^{2}\to1/2$ and $x$ is close to $1/\sqrt{2}$,
the integrand (a part from a factor $4/\pi$) behaves as (expanding
the argument of the square root up to second order around $x=1/\sqrt{2}$),
\begin{equation}
\frac{1}{2\sqrt{\left(x^{2}-y_{-}^{2}\right)\left(x^{2}-y_{+}^{2}\right)}}\simeq\frac{1}{2\sqrt{\frac{(1-2n)}{4}+2\left(x-\frac{1}{\sqrt{2}}\right)^{2}}}.
\end{equation}
This function can be integrated in any interval containing $x=1/\sqrt{2}$
leading to a logarithmic divergence. Specifically, using Eq.~(\ref{eq:divergence_identity}),
we obtain (without the factor $4/\pi$) 
\begin{equation}
-\frac{1}{2\sqrt{2}}\ln\left(\frac{1}{2}-n\right).
\end{equation}
Summing up both divergencies and multiplying by $4/\pi$ we obtain
the overall behavior for $n\to1/2^{-}$ 
\begin{equation}
P_{N_{2}}(n)=-\frac{3}{\sqrt{2}\pi}\ln\left(\frac{1}{2}-n\right)+O(1)\,.
\end{equation}
The behavior for $n\to1/2^{+}$ can be obtained in a similar way.
Exactly at $n=1/2$, there are non-integrable singularities at $x=0$,
$x=y_{-}$ and $x=y_{+}$. For the singularity around zero we can
approximate the integrand as 
\begin{equation}
\frac{1}{\sqrt{8}\sqrt{x^{2}+\left(n-\frac{1}{2}\right)}}.
\end{equation}
From this we obtain, for $n\to1/2^{+}$, integrating on any interval
$\left[0,\delta\right]$, with $\delta>0$, 
\begin{equation}
\int_{0}^{\delta}\frac{dx}{\sqrt{8}\sqrt{x^{2}+\left(n-\frac{1}{2}\right)}}=-\frac{1}{4\sqrt{2}}\ln\left(n-\frac{1}{2}\right)+O(1).
\end{equation}
For the singularities at $x=y_{\pm}$ we can approximate the integrand
(setting $y_{\pm}^{2}=1/2$ and $x$ close to $1/\sqrt{2}$) as 
\begin{equation}
\frac{1}{2\sqrt{2}\left|x-\frac{1}{\sqrt{2}}\right|}.
\end{equation}
Integrated around $y_{-}$ gives (for any $\delta<y_{-}$) 
\begin{equation}
\int_{\delta}^{y_{-}}\frac{dx}{2\sqrt{2}\left|x-\frac{1}{\sqrt{2}}\right|}=-\frac{1}{4\sqrt{2}}\ln\left(n-\frac{1}{2}\right)+O(1).
\end{equation}
The integral around $y_{+}$ gives the same result (for any $\delta>y_{+}$):
\begin{equation}
\int_{y_{+}}^{\delta}\frac{dx}{2\sqrt{2}\left|x-\frac{1}{\sqrt{2}}\right|}=-\frac{1}{4\sqrt{2}}\ln\left(n-\frac{1}{2}\right)+O(1).
\end{equation}
Summing up all three contributions and multiplying by $4/\pi$ we
obtain the same divergence as for $n\to1/2^{-}$. So, all
in all we obtain, for $n\to1/2$ 
\begin{equation}
P_{N_{2}}(n)=-\frac{3}{\sqrt{2}\pi}\ln\left|n-\frac{1}{2}\right|+O(1).
\end{equation}

\section{Absence of Van Hove singularities for larger Hilbert spaces}

\begin{figure}[ht]
  \centering
  \begin{minipage}{0.45\textwidth}
    \centering
    \includegraphics[width=\linewidth]{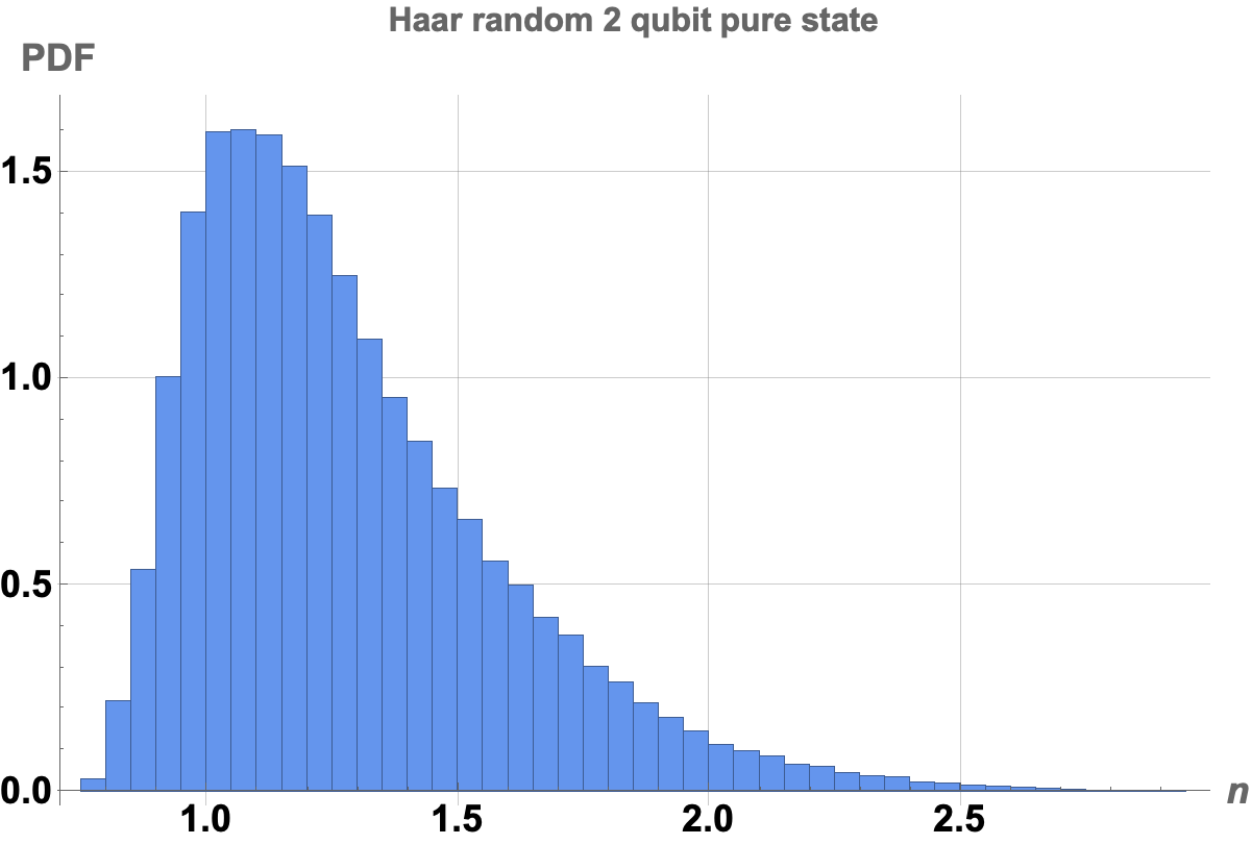}
  \end{minipage}
  \hfill
  \begin{minipage}{0.5\textwidth}
    \centering
    \includegraphics[width=\linewidth]{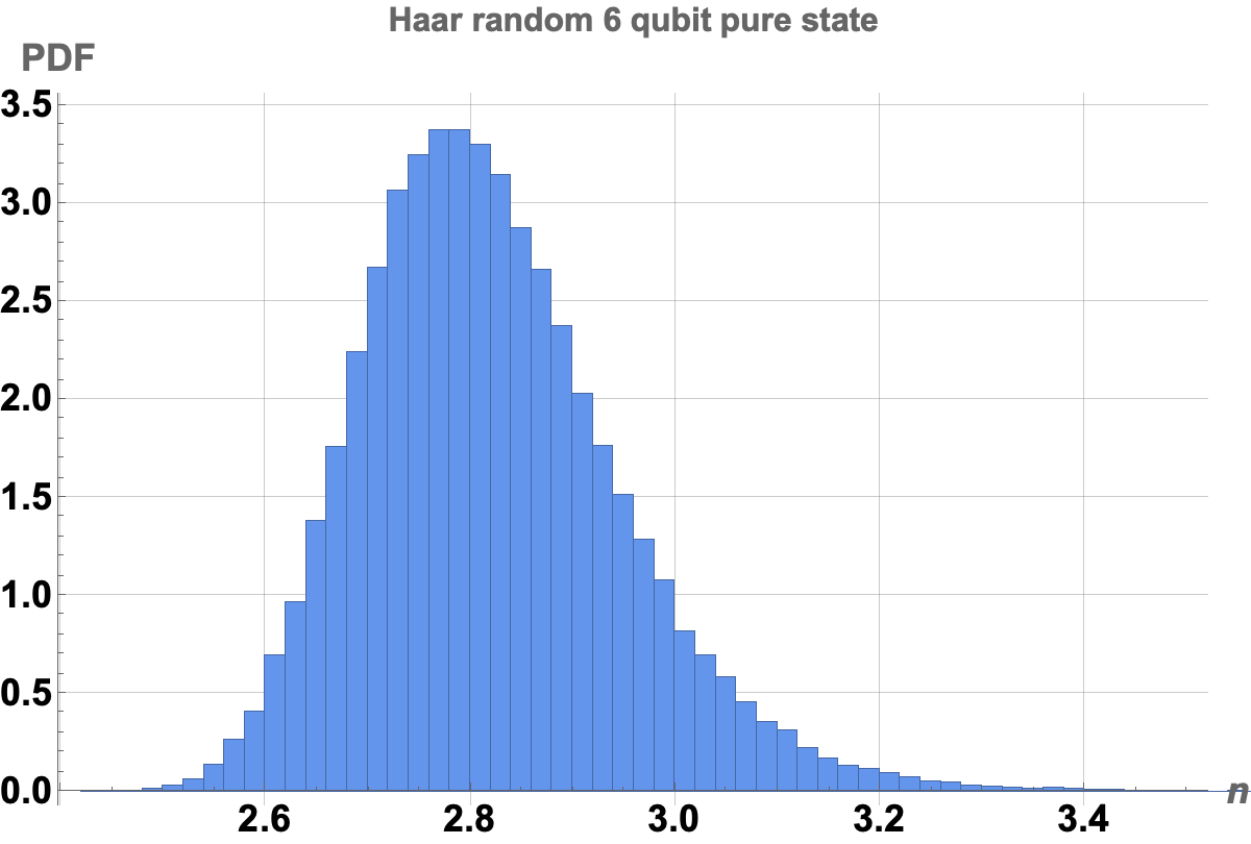}
    \label{fig:first}
  \end{minipage}
  \caption{Probability density function of $P_{N_2}(\textit{n})$ for 2 and 6 qubits extracted numerically for $N_{sample}=2 \times 10^5$ pure Haar random states.}
  \label{fig:Haar_no_sing}
\end{figure}

One can ask what happens in the case of larger Hilbert spaces such as
for qudit or multi-qubit systems. The construction of Eq.~(\ref{eq:Xi_2a})
can be generalized to any Hilbert space of dimension $d$. First, one writes
the general pure state in $\mathbb{C}P^{d-1}$ in terms of the Hurwitz
parametrization~\citep{Bengtsson_Zyczkowski_2006}, via $2d-2$ real
angles $\boldsymbol{\vartheta},\boldsymbol{\phi}$. If the system is a multi-qubit system, e.g.~$d=2^L$, one defines the vector $\boldsymbol{n}=\left(n_{1},\ldots,n_{d^{2}-1}\right)$ via 
\begin{equation}
n_{j}(\boldsymbol{\vartheta},\boldsymbol{\phi})=\mathrm{Tr}\!\left(P_{j}\,\psi\right),\quad j=1,\dots,d^{2}-1,
\end{equation}
for any Pauli string $P_{j}$ except the identity, that is $P_{j}\in\mathbb{P}_L \setminus \{\1\}$.
Then, for $\alpha\geq2$, one defines 
\begin{equation}
\Xi_{\alpha}(\boldsymbol{\vartheta},\boldsymbol{\phi})=\frac{1}{d}\left(1+\|\bm{n}(\boldsymbol{\vartheta},\boldsymbol{\phi})\|_{2\alpha}^{2\alpha}\right)=:\frac{1}{d}\left(1+N_{\alpha}(\boldsymbol{\vartheta},\boldsymbol{\phi})\right)\,,
\end{equation}
and we are led to study $P_{N_{\alpha}}(n)$ the Haar-induced probability
density  of $N_{\alpha}$. The problem becomes that of estimating
the DOS of a system with dispersion $N_{\alpha}(\boldsymbol{\vartheta},\boldsymbol{\phi})$
where the "Brillouin zone", here $\mathbb{C}P^{d-1}$, has dimension $D=2d-2$.
Now we borrow the result that, if the Hessian of $N_{\alpha}$ is
never identically vanishing (in other words, the critical points of
$N_{\alpha}$ are \emph{ordinary}), one does not have a Van Hove singularity
of the DOS for $D\ge3$~\cite{yuan_classification_2020}, meaning
$d\ge2.5$. Therefore, modulo unlikely accidental cancelation of
the Hessian, $d=2$ is the only dimension for which van Hove singularities
should be observed. Indeed, the absence of divergencies for $n>1$ is verified by our numerical simulations, see Fig.~\ref{fig:Haar_no_sing}.

\begin{figure}[ht]
  \centering
  \begin{minipage}{0.45\textwidth}
    \centering
    \includegraphics[width=\linewidth]{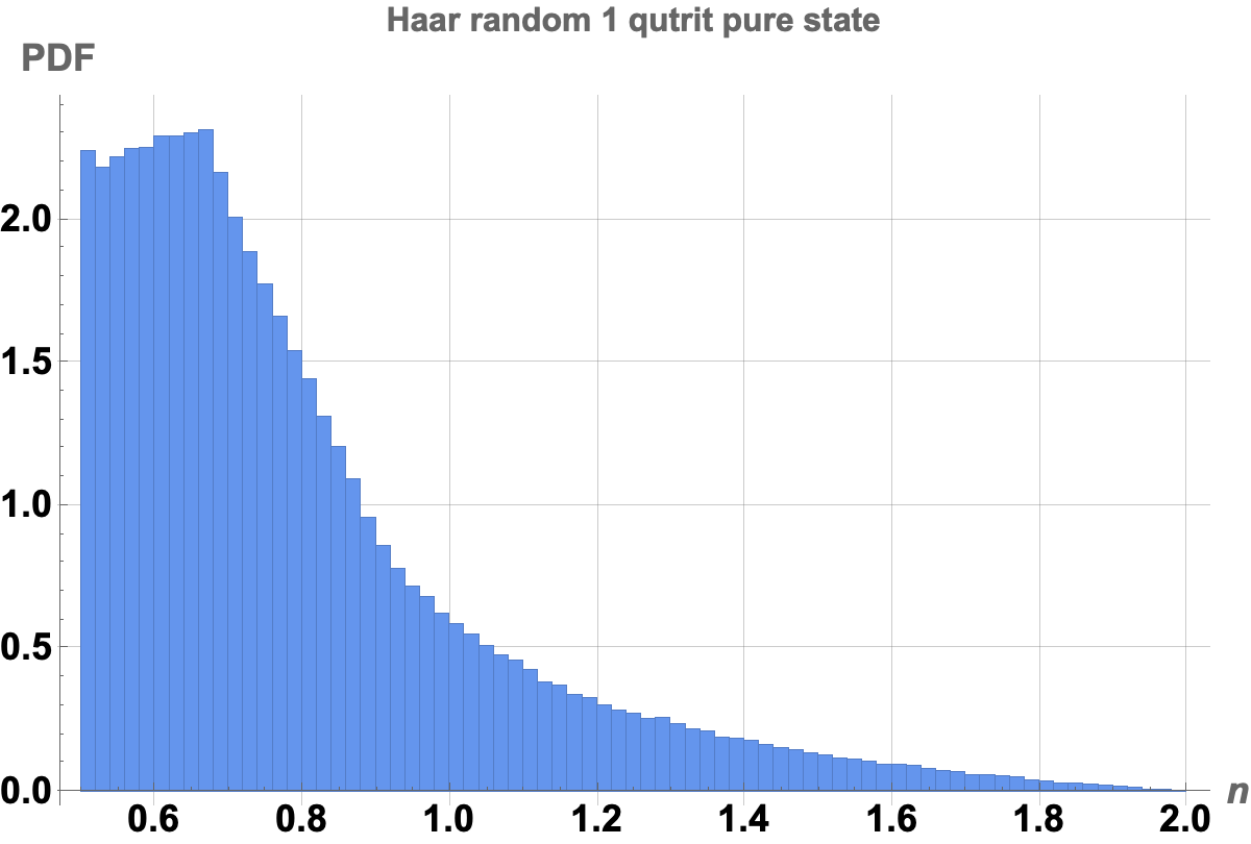}
  \end{minipage}
  \hfill
  \begin{minipage}{0.5\textwidth}
    \centering
    \includegraphics[width=\linewidth]{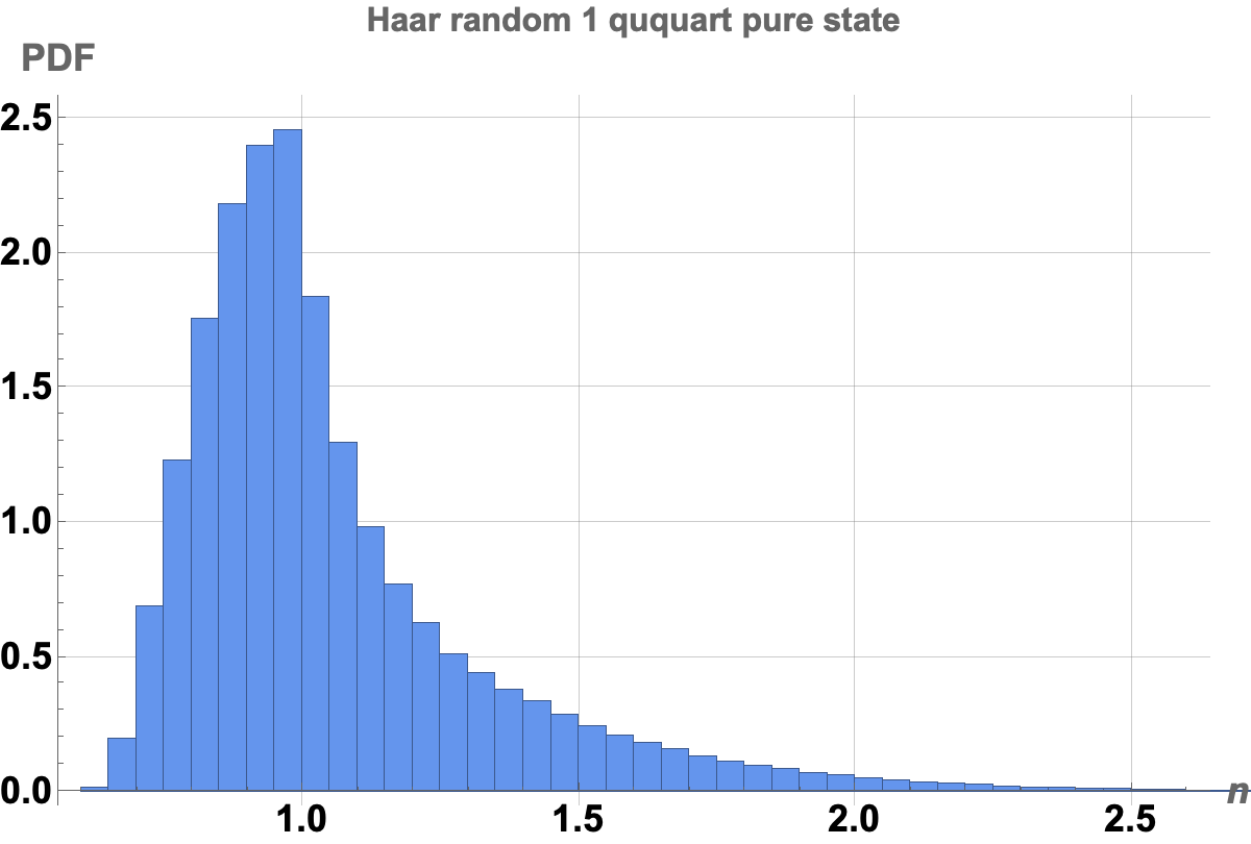}
    \label{fig:second}
  \end{minipage}
  \caption{Probability density function of $P_{N_2}(\textit{n})$ for one qutrit and ququart extracted numerically for $N_{sample}=4 \times 10^5$ pure Haar random states, where we used the generalization of SREs for the qudit case~\cite{Wang:2023uog}.}
  \label{fig:Haar_no_sing_qudit}
\end{figure}

In  case the system consists of one qudit of dimension $q$, one can extend the notion of the Pauli group to that of the Weyl-Heisenberg group, namely the set of displacement operators and integer powers of the phase factor $\omega$:
\begin{equation}
\mathcal{W}(q) = \{\, \omega^{s} D_{\textbf{a}} : s \in \mathbb{Z}_q,\; a \in \mathbb{Z}_q^2 \,\}.
\end{equation}
Thus, in analogy to the qubits case, the stabilizer purities in the case of qudit is defined as~\cite{Wang:2023uog}
\begin{equation}
    \Xi_{\alpha}(\ket{\psi})=\frac{1}{q} \sum_{D_\textbf{a} \in \mathcal{W}(d)} \abs{\text{Tr}(D_{\textbf{a}}\psi)}^{2 \alpha} ,
\end{equation}
and the SREs
\begin{equation}
    M_{\alpha}(\ket{\psi})=\frac{1}{1-\alpha} \ln \Xi_{\alpha}(\ket{\psi})\,.
\end{equation}
Finally, in this context $N_\alpha(\ket{\psi})= q \, \Xi_{\alpha}(\ket{\psi}) -1 $ (see Appendix \ref{App:qudits} for details), and after Hurwitz parametrization we are led formally to a similar problem as before. For what we have said, we do not expect divergencies in the PDF of $N_\alpha$ for $q \ge 2.5$, and at the same time  we expect a less smooth behaviour of the PDF the smaller the dimension. These features can be appreciated in our numerical simulations shown in Fig.~\ref{fig:Haar_no_sing_qudit}.

\section{General considerations on single qubit PDF's}

\label{sec:considerations}
For what we have said previously, borrowing results for the density
of states in solids, one may be led to conclude that van Hove singularities
in $d=2$ are present in the PDF, $P_{X}(x)$, of any quantity $X$
defined on the Bloch sphere, or,  in other words, that in $d=2$, divergencies
are the norm rather then the exception. However, we should remember
that the condition for logarithmic divergence in $d=2$, is that the
function $X(\vartheta,\phi)$ has a saddle point on the sphere. Moreover,
one should be careful that the PDF's that we are considering here
are defined on a curved base manifold, the (Bloch) sphere, as opposed
to a flat manifold as is the standard case for the density of states
in solids. In practice, one has to take into account the effect of the
term $\sin\left(\vartheta\right)$ in the measure in Eq.~\eqref{eq:PN_coarea}. See \cite{KENDON_2002} for some concrete examples in the case of different measures of entanglement.

To illustrate these subtleties, we consider then two other important
quantities that can be defined on the single-qubit space. The first one
is the expectation value $A=\langle\psi|\mathcal{A}|\psi\rangle$
of an observable $\mathcal{A}$. The PDF, $P_{A}(a)$, of an observable
$\mathcal{A}$, when $|\psi\rangle$ is uniformly distributed in $\mathbb{C}P^{d-1}$,
has been computed in~\cite{campos_venuti_probability_2013}. For $d=2$
the PDF has the following expression:
\begin{equation}
P_{A}(a)=\frac{1}{a_{2}-a_{1}}\1_{[a_{1},a_{2}]}(a)\,,\label{eq:PDF_observable}
\end{equation}
 where $a_{1}<a_{2}$ are the eigenvalues of $\mathcal{A}$, and $\1_{[a_{1},a_{2}]}(a)$
is the characteristic function of the interval $[a_{1},a_{2}]$. In
other words, $P_{A}(a)$ is the uniform distribution supported in $[a_{1},a_{2}]$.
The explicit form of $A(\vartheta,\phi)$ depends on the observable.
However, upon diagonalization, it can always be cast in the form 
\begin{equation}
U^{\dagger}\mathcal{A}U=\diag\left(\mathcal{A}\right)=\frac{a_{1}+a_{2}}{2}\1+\frac{a_{1}-a_{2}}{2}Z,
\end{equation}
 so that after a rotation, a shift, and a scale transformation, one
is led to consider $Z(\vartheta,\phi)=\cos\left(\vartheta\right)$
(or any other Pauli matrix for that matter). The problem becomes one-dimensional,
and according to the standard classification one would expect square
root singularities at the extrema of $\cos\left(\vartheta\right)$.
However, the measure is not $d\vartheta$ but rather $(1/2)\sin\left(\vartheta\right)d\vartheta$
and a single change of variable, $x=\cos\left(\vartheta\right)$,
leads to the result \eqref{eq:PDF_observable}. 

We now consider another quantity that can be defined on the Bloch
sphere. Namely the \emph{coherence} of the state $|\psi\rangle$ measured
using the computational basis as reference basis.
Using the $\ell^{1}$ norm, a measure of coherence in the
basis $\left\{ |i\rangle\right\} $ of the mixed state $\rho$,
is given by~\cite{baumgratz_quantifying_2014}
\begin{equation}
C(\rho)=\sum_{i\neq j}\left|\langle i|\rho|j\rangle\right|\,,
\end{equation}
which, for a single qubit, reduces to $C=2\left|\rho_{0,1}\right|$.
Using the standard parameterization of the sphere one obtains $C(\psi)=|\sin\vartheta|$,
which does not depend on $\phi$. Again, the problem is reduced to
computing a one dimensional DOS and one can expect a square root singularity
at the location where $C(\psi)$ is maximal or minimal. Indeed, in
this case, the singularity is not absorbed by the measure, and an explicit
calculation (see Appendix \ref{App:coherence}) gives
\begin{equation}
P_{C}(c)=\frac{c}{\sqrt{1-c^{2}}}\1_{[0,1]}(c)\,,    
\end{equation} 
which has a square root singularity at $c=1$, i.e.~at the maximum of $C$. 

This short discussion seems to indicate that, out of three physically interesting quantities that can be defined on a qubit, only the SRE shows a logarithmic singularity in its PDF. 
Moreover, the PDF of another measure of non-stabilizerness, the magic by trace distance, has been recently investigated numerically \cite{junior2025geometricanalysisstabilizerpolytope}. The results in \cite{junior2025geometricanalysisstabilizerpolytope} show a singularity also occurring at the $\ket{H}$-states \cite{albertojunior2025_privatecomm} and  are compatible with a logarithmic divergence at this critical value. This, if confirmed, indicates that the logarithimic divergence is a genuine feature of non-stabilizerness which does not depend on the particular monotone.

\section{Incompatibility deficit}
So far, we have studied for a single qubit the distribution of the stabilizer R\'{e}ny entropy and traced it back to that of the stabilizer purity. The stabilizer purity has a clear information-theoretic~\cite{leone2022StabilizerRenyiEntropy} and operational~\cite{PhysRevA.107.022429} interpretation. Remarkably, in the case of one-qubit systems, it has also a very clear interpretation in terms of fundamental quantum mechanics concepts. As is well known, one of the defining features of quantum mechanics lies in the existence of non-compatible observables and bases. This is, for example, demonstrated in  fundamental thought experiments such as the cascaded Stern-Gerlach experiments~\cite{sakurai2020modern}, where one shows that, for a two-level system, there are three maximally incompatible bases, typically referred to as the $X,Y,Z$ bases, corresponding to measurements characterized by statistics that go beyond classical results.
The fact that there are infinitely partially incompatible bases is usually overlooked as they show the same behavior, just in a lesser fashion. Now we argue that for one qubit, partial incompatibility is in fact directly related to non-stabilizerness.
Since compatible quantities commute, we  define the $\alpha-$incompatibility  as 
\begin{equation}
    \Gamma_\alpha (\psi ):= \sum_{j=1}^3 \norm{[\psi,\sigma_j]}_{2 \alpha}^{2 \alpha}=\frac{1}{2^{2 \alpha}}\sum_{j=1}^3 \norm{[\bm{n} \cdot \bm{\sigma},\sigma_j]}_{2 \alpha}^{2 \alpha}\,,
\label{eq:incomp}
\end{equation}
with  $\bm{\sigma}=\{X,Y,Z\}$ and $\norm{\bm{n}}_2=1$, where we identify $\psi= \frac{\id + \bm{n} \cdot \bm{\sigma}}{2}$. This is (proportional to) the average non compatibility of the single qubit state with respect to the Pauli operators $\sigma_j$ as measured by the $2\alpha-$norm of the commutator.
In order to compute the generic explicit formula, let us start by noticing that
\begin{equation}
    C_j:=[\bm{n} \cdot \bm{\sigma},\sigma_j]= \sum_{k=1}^3 n_k \,[\sigma_k ,\sigma_j]= 2 i \sum_{k=1}^3 n_k\, \varepsilon_{kjl}\, \sigma_l \equiv  2 i (\bm{n} \times e_j) \cdot \bm{\sigma},
\end{equation}
where $e_j$ is is the j-th standard basis vector and $v_j:= \bm{n} \times e_j$ is the vector whose components are $(v_j)_l=n_k\, \varepsilon_{kjl}$.
Observe that $\norm{v_j}^2_2=\norm{\bm{n} \times e_j}_2^2=\norm{\bm{n}}_2^2-n_j^2$.
Recall that the shatten norm is defined as $\norm{A}_{2 \alpha}^{2 \alpha}=\text{Tr}[(A^\dagger A)^\alpha]$, hence
\begin{equation}
    C_j^\dagger C_j= 4 (v_j \cdot \bm{\sigma})^2= 4 \norm{v_j}_2^{2} \id,
\end{equation}
which means
\begin{equation}
    \sum_{j=1}^3 \text{Tr}[( C_j^\dagger C_j)^\alpha]= \sum_{j=1}^3 2^{2\alpha+1} \norm{v_j}_2^{2 \alpha} = \sum_{j=1}^3 2^{2\alpha+1} (\norm{\bm{n}}_2^2-n_j^2)^{\alpha}\,.
\end{equation}
Finally, we have the following expression in terms of stabilizer purities for integer values of $\alpha$
\begin{equation}
    \Gamma_\alpha(\psi) = 2 \sum_{j=1}^3  (\norm{\bm{n}}_2^2-n_j^2)^{\alpha} = 2 \sum_{s=0}^\alpha \binom{\alpha}{s} (-1)^s \norm{\bm{n}}_{2 s}^{2s} = 2 \sum_{s=0}^\alpha \binom{\alpha}{s} (-1)^s (2\, \Xi_{ s}(\psi) - 1)\,.
\end{equation}
In particular, for $\alpha =1,2$, one obtains
\begin{equation}
    \Gamma_1(\psi)= 4 \norm{\bm{n}}_2^2=4\,,  \quad \Gamma_2(\psi)= 4 \Xi_2(\psi) =  4 (1-M^{\text{lin}}_2(\psi))\,.
    \label{eq:Gamma2}
\end{equation}
This result shows that the incompatibility  of order two {\em is} proportional to the stabilizer purity and the linear stabilizer entropy is an incompatibility deficit. Hence, the PDF of the two-incompatibility is readily obtained from the PDF of two-stabilizer purity, $\Xi_2$.  Moreover, Eq.~\eqref{eq:Gamma2} implies that stabilizer states host the maximum amount of incompatibility with respect the $X,Y,Z$ directions while non-stabilizerness is a deficit of such incompatibility. Notice that the $X,Y,Z$ are the same that define the notion of stabilizer states, that is, the Pauli algebra of one qubit.

A similar result can be obtained in the case of multiple-qubit systems, generalizing Eq.~(\ref{eq:incomp}) to 
\begin{equation}
\Gamma_\alpha(\psi)
=
\sum_{P\in\mathcal{P}_L}
\bigl\|[\psi,P]\bigr\|_{2\alpha}^{2\alpha},
\label{eq:Gamma-def}
\end{equation}
in which case one obtains (see Appendix \ref{sec:incomp_multiqubit})
\begin{equation}
     \Gamma_\alpha(\psi)= 2 d \sum_{s=0}^{\alpha}
\binom{\alpha}{s}(-1)^s \Xi_s(\psi),\quad \Gamma_2(\psi)=2d\bigl(d-2+\Xi_2(\psi)\bigr)= 2d\bigl(d-1-M^{\text{lin}}_2(\psi)\bigr),
\end{equation}
with $d=2^L$.

\section{Conclusions}

In this work, we have analyzed the Haar-induced probability distributions of measures of non-stabilizerness, specifically the stabilizer Rényi entropies, and uncovered a geometric structure underlying their behavior. Focusing on the single-qubit case, we have shown that the corresponding probability density functions exhibit non-analytic features in the form of logarithmic divergencies, directly analogous to Van Hove singularities in the density of states of condensed matter systems. 

By mapping the stabilizer purity to an effective dispersion relation defined on the Bloch sphere, we established a correspondence between the geometry of quantum states and the statistical structure of magic. The singularities emerge at saddle points of this dispersion, where the topology of the intersection between the $\ell_2$ and $\ell_{2\alpha}$ spheres changes, and are analytically captured by a logarithmic divergence of the probability density near the critical value $n_c = 2^{1-\alpha}$.
We derived the exact form of the probability density function for the case $\alpha = 2$, confirmed the logarithmic behavior through analytical asymptotics and numerical simulations, and computed the exact mean stabilizer Rényi entropy for one qubit. 
In practice, the observed divergencies in the magic density imply that, drawing states uniformly at random, most states will have magic similar to that of the $\ket{H}$-states -- a class of states useful for quantum computation.
Extending the analysis to higher-dimensional Hilbert spaces, we showed that such divergencies disappear for $d \ge 3$, in accordance with the general theory of densities of states in solids. 

Beyond the geometric interpretation of the probability densities, we also identified a direct physical meaning of the stabilizer purity for multiqubit systems: the linear stabilizer entropy quantifies an \emph{incompatibility deficit}, measuring the reduction in non-commutativity with respect to Pauli observables. This connects non-stabilizerness to one of the fundamental aspects of quantum mechanics, the incompatibility of measurements.

Taken together, these results demonstrate that the statistical structure of magic reflects the underlying geometry of quantum state space, and that singularities in its Haar distribution encode transitions between distinct geometric regimes of incompatibility. 

\section{Acknowledgments}
D.I. would like to acknowledge the Les Houches summer school 2025 on Exact Solvability and Quantum Information for the opportunity to work on this project while being there as well as B.~Jasser for a few comments on an early version of the manuscript. The authors thank Alberto B. P. Junior and coauthors in \cite{junior2025geometricanalysisstabilizerpolytope} for the useful exchange of emails. A.H.~acknowledges support from the PNRR MUR project PE0000023-NQSTI and the PNRR MUR project CN 00000013-ICSC.

\appendix
\section{Mean value of SRE for one qubit}
\label{App:Mean_SRE_1qubit}
Here we compute the exact expectation value of the stabilizer 2-Renyi entropy for one qubit
\begin{align}
    \mathbb{E}_\psi[M_2(\psi)]&= \int_{0}^{\ln \left( \frac{3}{2} \right)} dm\, m\, P_{M_2}(m)\\
    &= 2  \int_{0}^{\ln \left( \frac{3}{2} \right)} dm \,m \,e^{-m}\, P_{N_2}(2 e^{-m}-1)\\
    &= \int_{1/3}^1 dn \ln(\frac{2}{n+1}) P_{N_2}(n)\\
    &= \frac{4}{\pi} \int_0^1 dx \ln(\frac{2}{n+1}) \frac{\theta(A(x)-\abs{B(x)-4n})}{\sqrt{A(x)^2-(B(x)-4n)^2}}\\
    &= \frac{1}{\pi} \int_0^1 dx \int_{-A(x)}^{A(x)} du \frac{\ln \left( \frac{8}{B(x)-u+4} \right)}{\sqrt{A(x)^2-u^2}}\\
    &= \ln(8)-  \frac{1}{\pi} \int_0^1 dx \int_{-A(x)}^{A(x)} du \frac{\ln \left(B(x)-u+4 \right)}{\sqrt{A(x)^2-u^2}}\\
    &= \ln(8)-  \frac{1}{\pi} \int_0^1 dx \int_{-\pi/2}^{\pi/2} d\phi \ln \left(B(x)-A(x) \sin \phi+4 \right)\\
    &= \ln(8)-   \int_0^1 dx \ln \left(\frac{B(x)+4+\sqrt{(B(x)+4)^2-A(x)^2}}{2} \right)\\
    &= \int_0^1 dx \ln \left(\frac{16}{7 x^4-6 x^2+4 \sqrt{3 x^8-5 x^6+8 x^4-5 x^2+3}+7}\right) \simeq 0.228921,
\end{align}
using the following substitutions $n=2e^{-m}-1$, $u=B(x)-4 n$ and $u=A(x) \sin \phi$ with $A(x)=(1-x^2)^2$ and $B(x)=3 A(x)+4x^4$.

\section{Change of variables}
\label{App:Change_variables}
The aim of this section is to carry out a detailed calculation for Eq. \eqref{eq:SRE_SP_expantion_nc}.
When $n\to n_{c}$ we have that the PDF for $N_2$ reads
$$
P_{N_{\alpha}}(n)\propto-\ln\left|n-n_{c}\right|.
$$
Hence, given
\begin{equation}
P_{\Xi_{\alpha}}(\xi)=2P_{N_{\alpha}}(2\xi-1)\,,\quad P_{M_{\alpha}}(m)=2(\alpha-1)e^{(1-\alpha)m}P_{N_{\alpha}}(2e^{(1-\alpha)m}-1)\,.
\end{equation}
we can use the change of variables as follows
\begin{equation}
    \lim_{n\to n_c} P_{\Xi_{\alpha}}(\xi)=2 \lim_{n\to n_c} P_{N_{\alpha}}(2\xi-1) = 2 c \ln \abs{2\xi-1 - n_c}= 2 c \ln \abs{\xi-\frac{1+n_c}{2}} + 2 c \ln 2\,.
\end{equation}
where $c$ is a proportionality constant.
In the case of $M_2$ we have to consider $m_c= \frac{1}{1- \alpha} \ln \frac{1+n_c}{2}$ and $\delta m= m-m_c$ such as $\abs{\delta m}\ll 1$.
By considering the argument we have that 
\begin{equation}
    2e^{(1-\alpha)m}-1-n_c=2e^{(1-\alpha)m_c}(e^{(1-\alpha)\delta m}-1)= 2e^{(1-\alpha)m_c} (1- \alpha) \delta m + O(\delta m^2)\,.
\end{equation}
Thus
\begin{equation}
   P_{M_{\alpha}}(m) \propto-\ln\left|m-m_{c}\right|.
\end{equation}

\section{SRE for qudits}
\label{App:qudits}
When we deal with qudits instead of qubits we have to extend the notion of the Pauli group to that of the Weyl-Heisenberg group.
Consider a Hilbert space $\mathcal{H}_d$ of dimension $d$ with the orthonormal computational basis
$\{\lvert k\rangle : k \in \mathbb{Z}_d\}$, where $\mathbb{Z}_d = \{0,1,\ldots,d-1\}$ is the ring of integers modulo $d$.
We define two unitary operators, the \emph{shift} operator $X$ and the \emph{phase} (or \emph{clock}) operator $Z$ \footnote{In analogy to the Pauli matrices X and Z}, by
\begin{equation}
X\lvert k\rangle = \lvert k+1\rangle, \qquad
Z\lvert k\rangle = \omega^{k}\lvert k\rangle,
\end{equation}
where $\omega = e^{2\pi i / d}$ and all index arithmetic is taken modulo $d$.
These operators generate the basic algebraic structure of the Weyl--Heisenberg group and satisfy
\begin{equation}
X^d = Z^d =\id,
\end{equation}
and the fundamental commutation rule
\begin{equation}
X^k Z^\ell = \omega^{-k\ell}\, Z^\ell X^k,
\end{equation}
for all $k,\ell \in \mathbb{Z}_d$.
Given a pair of integers $\textbf{a}=(a_1,a_2)\in \mathbb{Z}_d^2$, we introduce the \emph{displacement operator}
\begin{equation}
D_{\textbf{a}} = \omega^{\tfrac{a_1 a_2}{2}} X^{a_1} Z^{a_2}.
\end{equation}
The prefactor involving $\tfrac{a_1 a_2}{2}$ is understood modulo $d$, and when $d$ is even one must adopt a consistent convention for the “half-integer” exponent.  
The operators $\{D_{\textbf{a}}\}_{\textbf{a} \in \mathbb{Z}_d^2}$ close under multiplication and obey the relations
\begin{equation}
D_{\textbf{a}}^\dagger = D_{- \textbf{a}},\quad D_{\textbf{a}} D_{\textbf{b}} = \omega^{\tfrac{1}{2}[\textbf{a},\textbf{b}]}\, D_{\textbf{a}+\textbf{b}},
\end{equation}
where the symplectic form on the discrete phase space $\mathbb{Z}_d^2$ is
$[\textbf{a},\textbf{b}] = a_1 b_2 - a_2 b_1$.
These operators form an orthogonal basis with respect to the Hilbert--Schmidt inner product, as expressed by
\begin{equation}
\mathrm{Tr}(D_{\textbf{a}} D_{\textbf{b}}^\dagger) = d\,\delta_{\textbf{a},\textbf{b}}.
\end{equation}
The full Weyl--Heisenberg group $\mathcal{W}(d)$ is the set of all products of displacement operators and integer powers of the phase factor $\omega$:
\begin{equation}
\mathcal{W}(d) = \{\, \omega^{s} D_{\textbf{a}} : s \in \mathbb{Z}_d,\; a \in \mathbb{Z}_d^2 \,\}.
\end{equation}
It contains $d^3$ elements, corresponding to the $d^2$ independent displacements and the additional $d$ possible global phases.
Global scalar phases, however, are physically irrelevant and form a central subgroup
\begin{equation}
\mathcal{Z} = \{\, \omega^{s} \1 : s \in \mathbb{Z}_d \,\}.
\end{equation}
Quotienting by this subgroup gives the \emph{phase-free Weyl--Heisenberg group}
\begin{equation}
\overline{\mathcal{W}}(d) = \mathcal{W}(d) / \mathcal{Z}.
\end{equation}
This quotient identifies all elements that differ by an overall $d$-th root of unity.  
The resulting group has order $d^2$ and corresponds to taking only those operators whose representative phases are $+1$.
We are then ready to the fine the stabilizer purities in the case of qudits as~\cite{Wang:2023uog}
\begin{equation}
    \Xi_{\alpha}(\ket{\psi})=\frac{1}{d} \sum_{\textbf{a}} \abs{\text{Tr}(D_{\textbf{a}}\psi)}^{2 \alpha} ,
\end{equation}
and the SREs
\begin{equation}
    M_{\alpha}(\ket{\psi})=\frac{1}{1-\alpha} \log_d \Xi_{\alpha}(\ket{\psi})\,.
\end{equation}
Finally, in this context $N_\alpha(\ket{\psi})= d\, \Xi_{\alpha}(\ket{\psi}) -1 $.

\section{Coherence}
\label{App:coherence}
In this section, we derive the PDF of the $\ell_1$ norm of coherence~\cite{baumgratz_quantifying_2014} of one qubit defined as
\begin{equation}
    C(\rho)= \sum_{i \neq j} \abs{\bra{i}\rho \ket{j}} = 2\abs{\rho_{0,1}},
\end{equation}
with fixed reference basis $\{\ket{i}\}$, which in this case we choose to be the Z basis.
Namely, in the case of one qubit pure state 
\begin{equation}
    \ket{\psi}=\cos\frac{\theta}{2} \ket{0} + e^{i \phi} \sin \frac{\theta}{2} \ket{1},\quad \theta \in [0,\pi]\, \text{and} \, \phi \in[0,2 \pi]\,,
\end{equation}
we have
\begin{equation}
    C(\ket{\psi}\bra{\psi})= \sin \theta\,.
\end{equation}
Hence,

\begin{align}
P_{C}(c) & =\int_{0}^{2\pi}\int_{0}^{\pi}\frac{\sin\theta d\theta d\phi}{4\pi}\delta(C(\theta)-c) = \frac{1}{2} \int_{0}^{\pi} \sin\theta\, d\theta \,\delta(C(\theta)-c)\\
 &= \frac{1}{2} \sum_{i=1}^2 \frac{\sin \theta_i}{\abs{\cos \theta_i}}=\frac{c}{\sqrt{1-c^2}}\,,
\end{align}
with $\sin \theta_i= c $ and $\cos \theta_i= \pm \sqrt{1-c^2}$.

\section{Incompatibility deficit for multi-qubits case}
\label{sec:incomp_multiqubit}
We have defined the partial incompatibility in Eq. \eqref{eq:Gamma-def} of order $\alpha \geq 1/2$ as
\begin{equation}
\Gamma_\alpha(\psi)
=
\sum_{P\in\mathcal{P}_L}
\bigl\|[\psi,P]\bigr\|_{2\alpha}^{2\alpha},
\end{equation}
where the Schatten $2\alpha$-norm is given by $\|A\|_{2\alpha}^{2\alpha}=\text{Tr}\!\left[(A^\dagger A)^\alpha\right]$.
Additionally, We recall the following identities, valid for pure states and Pauli
operators:
\begin{equation}
\psi^2=\psi,
\qquad
\psi^\dagger=\psi,
\qquad
P^2=\id,
\qquad
P^\dagger=P.
\label{eq:basic-identities}
\end{equation}
Using these properties, we can write
\begin{equation}
\bigl\|[\psi,P]\bigr\|_{2\alpha}^{2\alpha}
=
\text{Tr}\!\left(
\bigl([\psi,P]^\dagger[\psi,P]\bigr)^\alpha
\right).
\label{eq:commutator-norm}
\end{equation}
We now define the positive operator
\begin{align}
W
&:= [\psi,P]^\dagger[\psi,P]
= (\psi P - P\psi)^\dagger(\psi P - P\psi) \nonumber\\
&= (P\psi-\psi P)(\psi P - P\psi) \nonumber\\
&= P\psi^2P - P\psi P\psi - \psi P\psi P + \psi P^2\psi \nonumber\\
&= P\psi P + \psi - (P\psi P\psi + \psi P\psi P).
\label{eq:X-def}
\end{align}
Writing explicitly $\psi=|\psi\rangle\langle\psi|$ and defining
\begin{equation}
w_P := \langle\psi|P|\psi\rangle,
\label{eq:xP-def}
\end{equation}
we obtain
\begin{equation}
W
=
P\psi P + \psi - w_P(P\psi+\psi P).
\label{eq:X-final}
\end{equation}
The operator $W$ has support only on the two-dimensional subspace spanned by
$\{|\psi\rangle, P|\psi\rangle\}$.
It is easy to show that its nonzero eigenvalues are both $1-w_P^2$,
\begin{align}
W|\psi\rangle
&=
\bigl\{P\psi P+\psi-w_P(P\psi+\psi P)\bigr\}|\psi\rangle \nonumber\\
&=
P|\psi\rangle\langle\psi|P|\psi\rangle
+|\psi\rangle
-w_P\!\left(P|\psi\rangle
+ w_P|\psi\rangle\right) \nonumber\\
&=
(1-w_P^2)|\psi\rangle.
\end{align}
and
\begin{align}
W P|\psi\rangle
&=
\bigl\{P\psi P+\psi-w_P(P\psi+\psi P)\bigr\}P|\psi\rangle \nonumber\\
&=
P|\psi\rangle
+ w_P|\psi\rangle
- w_P\!\left(w_P P|\psi\rangle+|\psi\rangle\right) \nonumber\\
&=
(1-w_P^2)P|\psi\rangle.
\end{align}

All remaining eigenvalues vanish. Indeed, for any vector $|\chi\rangle$ satisfying
\begin{equation}
\langle\psi|\chi\rangle=0,
\qquad
\langle\psi|P|\chi\rangle=0,
\end{equation}
one finds
\begin{align}
W|\chi\rangle
&=
P|\psi\rangle\langle\psi|P|\chi\rangle
+|\psi\rangle\langle\psi|\chi\rangle
- w_P\!\left(
P|\psi\rangle\langle\psi|\chi\rangle
+|\psi\rangle\langle\psi|P|\chi\rangle
\right) \nonumber\\
&=0.
\end{align}
Therefore,
\begin{equation}
\text{Tr}[W^\alpha]
=
2(1-w_P^2)^\alpha.
\label{eq:trace-X}
\end{equation}
Substituting into Eq.~\eqref{eq:Gamma-def}, for integer values of $\alpha$ we obtain the following finite sum
\begin{equation}
    \Gamma_\alpha(\psi)=
2\sum_P (1-w_P^2)^\alpha =
2\sum_P\sum_{s=0}^{\alpha}
\binom{\alpha}{s}(-1)^s w_P^{2s} =
2d\sum_{s=0}^{\alpha}
\binom{\alpha}{s}(-1)^s \Xi_s(\psi),
\end{equation}
where $\Xi_s(\psi)$ denotes the stabilizer purity of order $s$.
In particular, for $\alpha=2$ we find
\begin{equation}
\Gamma_2(\psi)
=
2d\bigl(d-2+\Xi_2(\psi)\bigr).
\label{eq:Gamma2-final}
\end{equation}


\end{document}